\documentclass[aps,pre,superscriptaddress,twocolumn,10pt,nofootinbib]{revtex4-1}

\usepackage{xcolor,graphicx}
\usepackage{tikz, tikzscale}
\usetikzlibrary{shapes,arrows,positioning,decorations.pathreplacing,backgrounds}
\usepackage{fancybox}
\usepackage{amsmath,amssymb,gensymb}
\usepackage[colorlinks=true,linkcolor=blue,anchorcolor=black,citecolor=red,
					filecolor=cyan,menucolor=red,runcolor=cyan,urlcolor=magenta]{hyperref}

\def\wid{0.95}

\DeclareMathOperator{\upd}{d\!}

\usepackage{amsfonts}

\newcommand{\be}{\begin{equation}}
\newcommand{\ee}{\end{equation}}
\newcommand{\bea}{\begin{eqnarray}}
\newcommand{\eea}{\end{eqnarray}}

\begin{document}

\title{Active sorting of particles as an illustration of the Gibbs mixing paradox}

\author{Cato Sandford}
\affiliation{Center for Soft Matter Research and Department of Physics, New York University, 726 Broadway, New York, NY 10003, USA}
\author{Daniel Seeto}
\affiliation{Stern School of Business, New York University, 44 West 4\textsuperscript{th} Street, New York, NY 10012, USA}
\author{Alexander Y. Grosberg}
\affiliation{Center for Soft Matter Research and Department of Physics, New York University, 726 Broadway, New York, NY 10003, USA}

\date{\today}


\begin{abstract}

The Gibbs Mixing Paradox is a conceptual touchstone for understanding mixtures in statistical mechanics. While debates over the theoretical subtleties of particle distinguishability continue to this day, we seek to extend the discussion in another direction by considering devices which can only distinguish particles with limited accuracy. We introduce two illustrative models of sorting devices which are designed to separate a binary mixture, but which sometimes make mistakes. In the first model, discrimination between particle types is passive and sorting is driven, while the second model is based on an active proofreading network, where both discrimination and sorting have a tunable active component. We show that the performance of these devices may be enhanced out of equilibrium, and we further probe how the quality of particle sorting is maintained by trade-offs between the time taken and the energy dissipated.  Considering these examples, we demonstrate how increasing the similarity between particles gradually increases the work required to sort them, eliminating the paradox, while preserving the limits imposed by standard equilibrium statistical mechanics.

\end{abstract}


\maketitle

\section{Introduction}
\label{sec:intro}

\subsection{The goal of this article}

The goal of this article is to look from a unified point of view at two related issues: the famous Gibbs Mixing Paradox in equilibrium statistical mechanics, and the more practical issue of particle sorting, relevant to a number of non-equilibrium processes in biological cells.

\begin{figure}
	\centering
	\includegraphics[width=0.45\textwidth, angle=0]{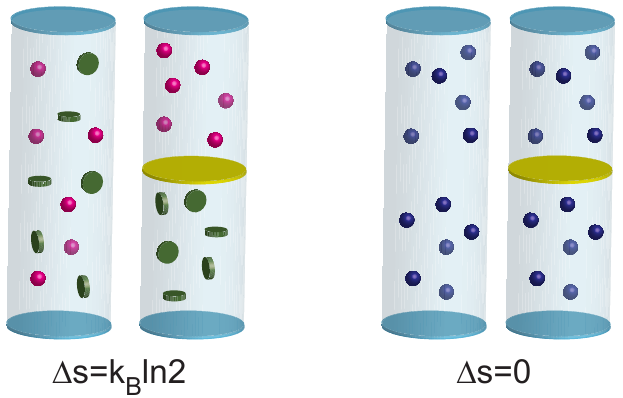}
	\caption{For the system on the left containing two different particle species, the entropy change upon separating them into two half-volumes is $\Delta s = -k_{\rm B} \ln 2$ per particle.  For the system on the right with identical particles, the entropy change is obviously zero, $\Delta s = 0$.}
	\label{fig:CompressBox_0}
\end{figure}

The Gibbs Mixing Paradox concerns the curious fact that, when separating a mixture of ideal gases into two volumes, the change in entropy does not depend on the nature of the gases (and similarly for the reverse process when two different gases mix into a single volume).  This is illustrated in Fig.~\ref{fig:CompressBox_0}, where gases are depicted as particles of different colors and shapes: the mixing entropy for two equal sub-volumes is $\Delta s = k_{\rm B} \ln 2$ per particle if the gases are different, and $\Delta s = 0$ if they are identical.  But this implies that if we make the two particle types \textit{arbitrarily similar} to one another ---though not identical---, we would continue observe the very same $\Delta s = k_{\rm B} \ln 2$, right up until the particles are indeed identical and the entropy change drops to zero. Gibbs himself noted this apparent discontinuity \cite{Gibbs1,Gibbs2,Gib02}, though he did not call it a paradox.%
\footnote{This is consistent with the fact that the issue is not even mentioned in some of the best statistical physics textbooks, eg \cite{LandauLifshitz_StatisticalPhysics}.}

Before we elaborate upon this (in section~\ref{sec:Idealisations}), it is worth emphasising that the physics of mixing and de-mixing is relevant to a variety of real-world phenomena. Biological cells, for instance, employ a suite of mechanisms to transport target molecules across membranes in order to maintain a given purity for the cellular environment, or maintain electrochemical potential differences.
While some elements of these processes may be \textit{passive} (ie, the system approaches an equilibrium state determined by the underlying free energy landscape), living systems are essentially non-equilibrium. Past work has therefore sought to explain and quantify non-equilibrium contributions to particle recognition and transport in several specific circumstances \cite{Hop74,Nin75,W+N03}. This paper approaches the same topics from a more basic statistical physics standpoint, employing toy models to elucidate essential characteristics of particle-sorting phenomena. As we shall discuss in the following section, a key consideration for us is the fact that any real device tasked with identifying and sorting different types particles will occasionally make mistakes, which influences thermodynamic outcomes, such as the work done or the entropy change from a sorting process.

\subsection{Idealisations in Statistical Mechanics}
\label{sec:Idealisations}

While the resolution to the Gibbs Mixing Paradox is widely held to be quantum mechanical \cite{Sa+06,Mat08,Kar07,For13,Da+11,Stu03,Tuc10,Hua87,Rei98} (see also further discussion in \cite{Pau73,B-N07,Swe15}), this point of view conspicuously fails to account for the usefulness and accuracy of statistical mechanics in the context of classical systems, such as colloids or proteins. Such particles are not strictly identical, yet may still be treated as such \cite{Les80} -- a fact compellingly emphasised in recent works \cite{Fre14,C+M15}.  In this context, the unsettling discontinuity of the Gibbs Paradox, between distinguishable and indistinguishable particles, arises from the temptation to regard entropy as an absolute quantifier of a system's properties, rather than a function of the chosen {macrostate} \cite{Kam84,Blumenfeld,Jay96,T+C02}. In fact, particle identity is merely one component of the macrostate, to be included whenever necessary. This is especially clear for classical particles, which are always distinguishable in principle \cite{Fon64,V+D11}.

Macrostates, the dichotomy between distinguishable and indistinguishable particles, and even entropy itself, are some of the idealisations employed by equilibrium statistical mechanics. It is these idealisations that give rise to the Gibbs Paradox, and so in this work, we ask to what extent they are applicable in case of real thermal particles (like colloids or proteins) which may be so similar that distinguishing them, although possible in principle, may be {difficult} for a given apparatus and prone to {errors}.  We may then consider a continuum between particles being distinguishable and indistinguishable.

\begin{figure}[ht]
	\centering
	\includegraphics[width=0.45\textwidth, angle=0]{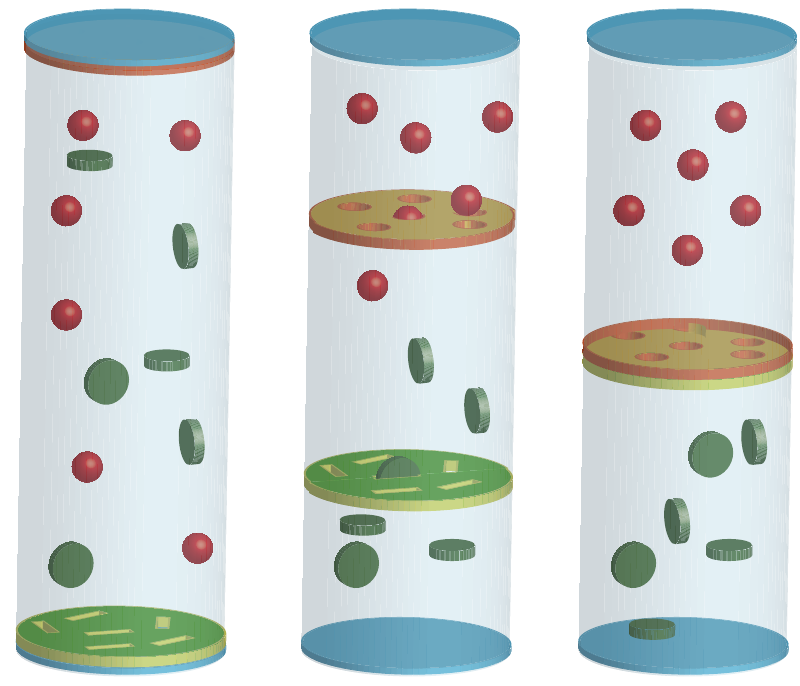}
	\caption{From left to right: A mixture of gases is sorted into two volumes by compressing with special pistons. The process requires at least $k_{\rm B}T\ln2$ of work per particle. However, if the pistons do not distinguish the two types of particles, no work is performed.}
	\label{fig:CompressBox}
\end{figure}

For example, suppose we wish to purify a binary mixture of gases into two volumes, as in Fig.~\ref{fig:CompressBox}.  To achieve this, we must have some way of determining the identity of each gas particle. In the cartoon, this is done by sieves on each of the two pistons, which in an ideal scenario are perfectly permeable for one particle species and impermeable for the other.  Each piston therefore compresses only one of the two mixed gases and performs work of at least $k_{\rm B}T \ln2$ per particle (assuming the system is in contact with a thermostat of temperature $T$).  This situation is ideal in the sense that the macrostate implies that particles can be distinguished, that distinguishing them does not require any work, and that entropy is a good proxy for the work done to separate them. It is under these circumstances that we encounter the ``paradox''.

\subsection{The plan of this article}

An important feature of the sorting process depicted in Fig.~\ref{fig:CompressBox} is that particles in the ``right'' sub-container have the same energy as in the ``wrong'' one; in other words, sorting here is purely kinetic. Another example of purely kinetic sorting, which we shall investigate in some detail, is more reminiscent of a Maxwell demon (sometimes called a ``concentration demon'' \cite{Phillips_book}). Using the device depicted in Fig.~\ref{fig:KineticBox}, we treat particle distinguishability not as a ``yes-no'' binary, but as a continuous parameter which encapsulates the accuracy of an {imperfect} sorting device.  We investigate in section~\ref{sec:KineticSorting} the consequences of the device's mistakes, for instance the impact on the work required to achieve a given level of purity, and how the device's performance may be improved through the introduction of energy-consuming discrimination steps.  In section~\ref{sec:HNsorter} we consider a fundamentally different situation in which particles have some pre-existing energetic preference for one of the boxes, such that a certain level of sorting will be achieved even in equilibrium. We then show how an active process, similar to kinetic proofreading \cite{Hop74,Nin75}, may be employed to improve the sorting quality.

In the end we show how the practically and biologically important issue of particle sorting sheds light on the Gibbs paradox, showing that it arises from unwarranted generalizations of some of the idealized concepts in equilibrium statistical mechanics.

\section{Purely Kinetic Sorting Without Energetic Preference}
\label{sec:KineticSorting}

\subsection{Sorting Model}
\label{sec:KineticSortingIntro}

Our ``purely kinetic'' model is broadly similar to the illustration in Fig.~\ref{fig:CompressBox}, in that sorting depends on an external driving. We shall first introduce a very simple version of the model with only passive discrimination, and then expand it to be slightly more complicated and versatile. Finally, we incorporate \textit{active} discrimination and describe how it can improve the sorting performance.

\begin{figure}[hb]
	\centering
	\includegraphics[width=0.8\linewidth]{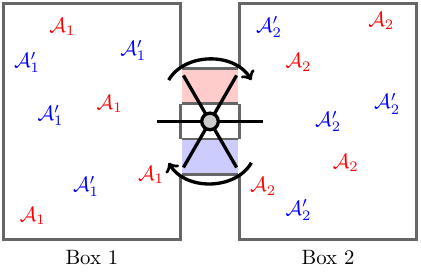}
	\caption{Schematic of a sorting device. Two types of particles are contained in two volumes connected by two channels. These (entirely passive) channels preferentially pass one type of particle: the upper channel prefers $\mathcal A$ and the lower channel prefers $\mathcal A^\prime$. Meanwhile, the central turbine indiscriminately pushes particles in the upper channel toward box~2, and those in the lower channel toward box~1 -- an action which costs some amount of free energy. The combined action of the channels and the turbine is to reach a sorted steady state, whose purity depends on the relative probability of errors, $\eta$.}
	\label{fig:KineticBox}
\end{figure}

Consider a system with two species of particles, $\mathcal A$ and $\mathcal A^\prime$, initially distributed evenly between two equal volumes, labelled 1 and 2. Notation-wise, we denote a particle of $\mathcal A$ in box 1 as $\mathcal A_1$, etc.  We use the same symbols also for the numbers of corresponding particles, so that $\mathcal{A}_1(t)$ is the number of $\mathcal{A}$ particles in box 1 at time $t$, etc.  The sorting device is sketched in Fig.~\ref{fig:KineticBox}.

Sorting proceeds through the combination of two processes. First, particles are \textit{distinguished} (passively) by the two coloured channels, which allow passage to particles of a particular type, so $\mathcal A$-type particles can move only through the upper channel, and $\mathcal A^\prime$-type particles can move only through the lower channel (we might imagine that the particles are of different shapes, as in figure \ref{fig:CompressBox}, and place a sieve at each entrance of each channel). The second process is \textit{driving}, which is performed by the turbine. Driving is insensitive to the particles' type, but induces them to flow in a particular direction -- those in the upper channel flow toward box 2, and those in the lower channel toward box 1.
\footnote{While the devices in Figs.~\ref{fig:CompressBox} and~\ref{fig:KineticBox} operate on the same principle, the virtue of the latter is that it can be run continuously. Thus, when we introduce mistakes in the discrimination step (so that the channels ``leak''), the device will still reach a non-trivial steady state. Fig.~\ref{fig:CompressBox}, on the other hand, would after a long time return to a maximum entropy equilibrium unless the pistons are made to continually recompress the gas.}

\subsection{An Elementary Example}
\label{sec:KineticSortingSimple}

To introduce the problem, let us first consider its most elementary version.  Suppose the sorters at the entrances of both channels make no mistakes, and pass through only their ``own'' particles with rate constant $K$.  Then
\begin{align}\begin{split}
	\dot{\mathcal{A}}_1(t) & = - K \mathcal{A}_1(t) \ , \ \  \mathcal{A}_1(t)+\mathcal{A}_2(t)=N \\
	\dot{\mathcal{A}^{\prime}}_2(t) & = - K \mathcal{A}^{\prime}_2(t) \ , \ \  \mathcal{A}^{\prime}_1(t)+\mathcal{A}^{\prime}_2(t)=N \ ,
\end{split} \label{eq:simplest_no_mistakes} \end{align}
where $N$ is the total number of particles of either type.  This of course results in $\mathcal{A}_1(t) = \mathcal{A}^{\prime}_2(t) = \frac{N}{2}e^{-Kt}$, and $\mathcal{A}_2(t) = \mathcal{A}^{\prime}_1(t) = N-\frac{N}{2}e^{-Kt}$.

How much work does the turbine perform? To transfer one $\mathcal{A}$ particle from box 1 to box 2, one needs to perform work which is at least equal to the difference of chemical potentials between two boxes, $\Delta \mu = k_{\rm B} T \ln \frac{\mathcal{A}_2(t)}{\mathcal{A}_1(t)}$ (and similarly for primed particles).  Since the current through either channel is $J(t) = - \dot{\mathcal{A}}_1(t)$, the minimal work to complete the separation is
\begin{align} W = \int_0^{\infty} J(t) \Delta \mu(t) \upd t =  k_{\rm B} T \int_0^{\infty} -\dot{\mathcal{A}}_1(t) \ln \frac{\mathcal{A}_2(t)}{\mathcal{A}_1(t)} \upd t \ . \label{eq:work_no_mistakes} \end{align}
Evaluating this integral yields the familiar answer of $k_{\rm B} T \ln 2$ per particle. As already mentioned, the $W$ given by formula~(\ref{eq:work_no_mistakes}) is the \textit{minimal} amount of necessary work: it cannot be achieved in practice, and one can only approach it if the process is run so slowly (i.e., if $K$ is so small) that the gases in each volume are completely equilibrated at all times. (It would, of course, also require the avoidance of dissipation in the turbine and everywhere else.) Our point is to emphasize that the required minimal work of an idealized, dynamical, error-free process is indeed limited by the equilibrium entropy, in perfect agreement with thermodynamics.

Now, while the channels are \textit{intended} to allow passage for one type of particle only, let us imagine that the other type may still leak through. We introduce the relative probability of such an error $\eta$ (to be defined more precisely below), and note that $\eta$ is a proxy for the distinguishability of the particle types: $\eta=0$ corresponds to zero mistakes and perfect distinguishability, while $\eta=1$ corresponds to perfect indistinguishability.

The possibility of a particle mistakenly going through the wrong channel modifies kinetic equations~(\ref{eq:simplest_no_mistakes}):
\begin{align}\begin{split}
	\dot{\mathcal{A}}_1(t) & = - K \mathcal{A}_1(t) + K \eta \mathcal{A}_2(t) \ , \ \  \mathcal{A}_1(t)+\mathcal{A}_2(t)=N \\
	\dot{\mathcal{A}^{\prime}}_2(t) & = - K \mathcal{A}^{\prime}_2(t) + K \eta \mathcal{A}^{\prime}_1(t) \ , \ \  \mathcal{A}^{\prime}_1(t)+\mathcal{A}^{\prime}_2(t)=N \ ,
\end{split} \label{eq:simplest_with_mistakes} \end{align}
leading to
\begin{align}
	\frac{\mathcal{A}_1(t)}{N} = \frac{\eta}{1+\eta} + \left[ \frac{1}{2} - \frac{\eta}{1+\eta} \right] e^{-K(1+\eta)t} \ ,
\end{align}
and similarly for primed particles (henceforth we will not repeat symmetrical-looking equations for both species of particles).  As expected, when mistakes are possible, each box remains contaminated with incorrect particles even after infinite time devoted to sorting. {This is our first glimpse of a ``softening'' of the Gibbs Paradox, where values of $\eta$ intermediate between zero and unity leads to values of the system's \textit{minimal possible} entropy change intermediate between $-N k_{\rm B} \ln 2$ and $0$ respectively.}

Although by the structure of kinetic equations~(\ref{eq:simplest_with_mistakes}) $\eta$ masquerades itself as an equilibrium constant between states $\mathcal{A}_1$ and $\mathcal{A}_2$, we prefer to think of it differently. Following on from the previous setup, we assume that transport through the upper channel is always to the right, and transport through the lower channel is always to the left, so that the detailed balance is violated.  In this sense, we treat $\eta$ as a quantifier of the ``partial distinguishability'' of particles.  In such interpretation, it has no direct thermodynamic meaning -- and it should not, because it is a kinetic (transport) property.

The issue of work in this case becomes interesting.  As long as $\eta$ is very small, the initial kinetics of separation is almost the same as in the mistake-free system, and the minimal amount of work performed in one channel $W(t) = k_{\rm B}T \int_0^{t} K \mathcal{A}_1(t^{\prime}) \ln \frac{\mathcal{A}_2(t^{\prime})}{\mathcal{A}_1(t^{\prime})} \upd t^{\prime}$ increases initially with time $t$ following very nearly the same schedule.  If $\eta$ is really very small, then the work approaches $k_{\rm B}T \ln 2$ per particle.

However, for times larger than $\ln\left({1}/{\eta}\right)/K$, mistakes start to take their toll: some particles return to their original box, going down the gradient of chemical potential, and have to be transported back a second time.  Thus the device has to perform work at a steady rate forever to maintain a bounded level of purity%
\footnote{This assumes, of course, that when a particle leaks through the wrong channel it does not return the corresponding energy to the turbine.}
(see also related theoretical work in Ref. \cite{HZE17}).

This model illustrates some important ideas about particle sorting. However, it lacks the flexibility to introduce active control of errors. Therefore, we now introduce a more sophisticated model.


\subsection{Kinetic Sorting with Passive Discrimination}

Starting from this section, we drop $k_{\rm B}T$ from all equations, assuming all energies measured in the units of $k_{\rm B}T$.

\begin{figure*}[ht]
	\centering
	\includegraphics{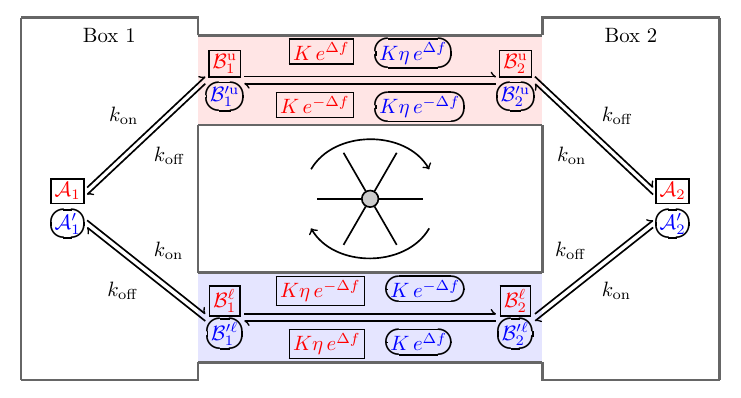}
	\caption{A kinetic network representing the setup in Fig.~\ref{fig:KineticBox}. Quantities relating to particle type $\mathcal A$ are coloured red and enclosed in rectangular boxes, while quantities relating to $\mathcal A^\prime$ are coloured blue and enclosed in curved boxes. An $\mathcal A$ particle in box~1 or~2 is denoted $\mathcal A_1$ and $\mathcal A_2$ respectively. When a particle $\mathcal A$ enters a channel it is denoted $\mathcal B$, with indices indicating which side of which channel it's in. With no driving, the upper channel passes $\mathcal A$ particles with rate $K$ in both directions, and $\mathcal A^\prime$ particles with rate $\eta K$ (where $0\leq\eta\leq1$ is the relative probability that a channel accepts the wrong particle). In the lower channel, these rates are reversed. Driving is parameterised by the factor $e^{\Delta f}\geq 1$ (as mentioned in the main text, we assume energy is measured in units of $k_{\rm B}T$). While it is not strictly necessary to have driving in both directions, it's convenient to make the networks as symmetical as possible.}
	\label{fig:KineticNet}
\end{figure*}

The updated model is illustrated in Fig.~\ref{fig:KineticNet}. Once again, the two passive channels allow passage of the different particles with rates $K$ and $\eta K\ll K$, while the turbine pushes particles in a direction which depends solely on which channel they're in and {not} on their type.

In contrast to the previous section, we shall assume here that the turbine works at a {constant} rate, expending a free energy of $\Delta f$ for every particle it pushes through the channel in the direction of its drive, and receiving $\Delta f$ for each particle that pushes it in the opposite direction. This is true independently of the state of the system. 
Using $\Delta f$ rather than $\Delta \mu$ will simplify our calculations a little. It also means that even when the channels perfectly distinguish the two particle types ($\eta=0$), there will be imperfect sorting, since particles are allowed to make transitions away from their target box as long as they pay $\Delta f$ to the turbine. While this consideration changes the details of the results, it doesn't affect the overall thrust of our findings.

We translate the Fig.~\ref{fig:KineticBox} setup into the reaction network in Fig.~\ref{fig:KineticNet}, which shows the paths a particle of type $\mathcal A$ or $\mathcal A^\prime$ may take to transition between the boxes. The states $\mathcal B$ and $\mathcal B^\prime$ denote a particle which has entered one of the channels, with indices ${\rm u,\ell}$ denoting whether it is in the upper or lower channel, and indices $1,2$ denoting which side of the channel it's in. (Although this may seem crudely reminiscent of some specific biological systems, our emphasis here is on the statistical physics aspect of things.)

As in the previous section, transitions between states are governed by kinetic rates.  The diffusion-controlled rates $k_{\rm on}$ and $k_{\rm off}$ represent respectively the rate at which a particle in one of the boxes enters a channel, and vice-versa. The rates $K$ and $K\eta$ represent transitions within the two channels when there is no assistance from the turbine (that is, when $\Delta f=0$).

\begin{figure}[ht]
	\centering
	\includegraphics[width=\wid\linewidth]{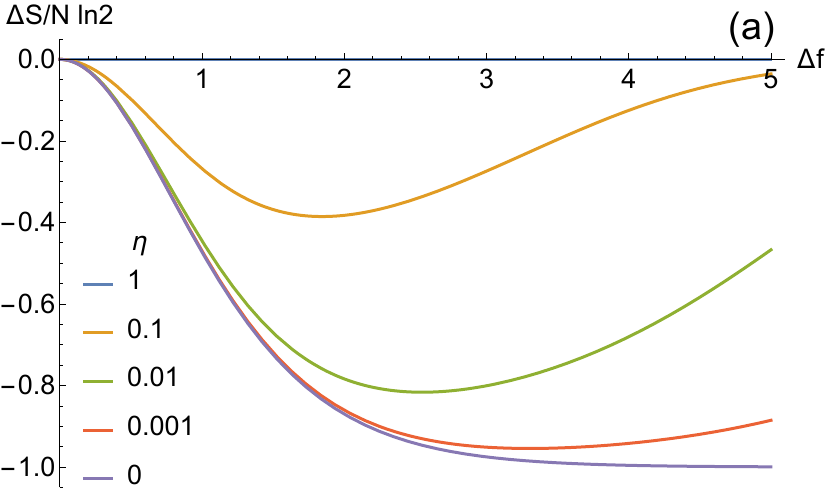}\\[1em]
	\includegraphics[width=\wid\linewidth]{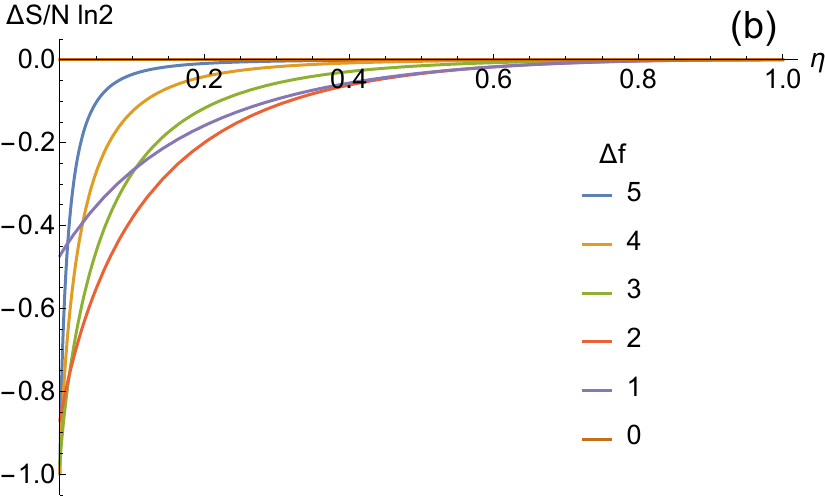}
	\caption{For the kinetic sorter in Fig.~\ref{fig:KineticNet}, we plot the sorting quality, $\Delta S$, against turbine driving $\Delta f$ and the error parameter $\eta$. $\Delta S$ is the entropy change from the unsorted initial state to the partially-sorted steady state, so more negative values correspond to better sorting. Other parameters are fixed to be $k_{\rm on}=10^{-3}k_{\rm off}$ and $K=0.1k_{\rm off}$. {(a) $\Delta S(\Delta f)$ for several values of $\eta$. As expected, $\eta=1$ (indistinguishable particles) leads to zero change in the state of the system, while $\eta=0$ leads to complete segregation for sufficiently high driving. (b) $\Delta S(\eta)$ for multiple values of $\Delta f$. The same limits apply for high and low $\eta$, and we clearly see how intermediate values of distinguishability influence the final sorting quality.}}
	\label{fig:KineticSortingEntropy}
\end{figure}

Representing this network as a set of linear, coupled, ordinary differential equations, we may easily calculate the steady state, and hence the sorting quality, as a function of $\Delta f$ and $\eta$. This is plotted in Fig.~\ref{fig:KineticSortingEntropy}, where the sorting quality is parameterised by the entropy change $\Delta S$ between the unsorted initial state and the maximally (but imperfectly) sorted steady state (see Appendix~\ref{app:Entropy} for the explicit calculation of the entropy). From panel (a) we see, as expected, that sorting is best when $\eta$ is small, and is poor when the driving $\Delta f$ is either very low or very high.%
\footnote{When the driving is low, there is insufficient incentive for particles to switch to the correct box. When the driving is high, there's so much incentive for particles to switch that both types of particles circulate near-freely between the boxes.}
This is confirmed in panel (b), which furthermore illustrates how the Gibbs Paradox's discontinuity is softened in the present scheme when the driving is sufficiently high (in this case when $\Delta f\gtrsim 2$).

\subsection{Kinetic Sorting with Active Discrimination}
\label{sec:KineticSortingActive}

For a given error $\eta$ and driving $\Delta f$, we seek to invest work from some energy reservoir to improve the sorting quality. The obvious solution would be to introduce an additional \textit{active sorting} mechanism, to shuttle particles in one direction or another depending on their type. This will be explored in more depth in section~\ref{sec:HNsorter}; for now we shall focus on a slightly subtler mechanism of \textit{active discrimination}. By this we mean an active process which is sensitive to particle's \textit{type}, but is agnostic about which box it should be assigned to. In other words, the active discrimination adds to the existing structure of the Fig.~\ref{fig:KineticNet} network in a way that is symmetrical with respect to the box numbers, but asymmetrical with respect to the particle types.

To give a further hint of what this means, recall that the sorting device introduced in Figs.~\ref{fig:KineticBox} and~\ref{fig:KineticNet} works because the flow of particle type $\mathcal A$ is slower in the lower channel than in the upper channel (and vice-versa for $\mathcal A^\prime$). Our aim is simply to enhance this discrepancy through a modification of the existing kinetics.

\begin{figure*}[ht]
	\centering
	\includegraphics[width=0.8\linewidth]{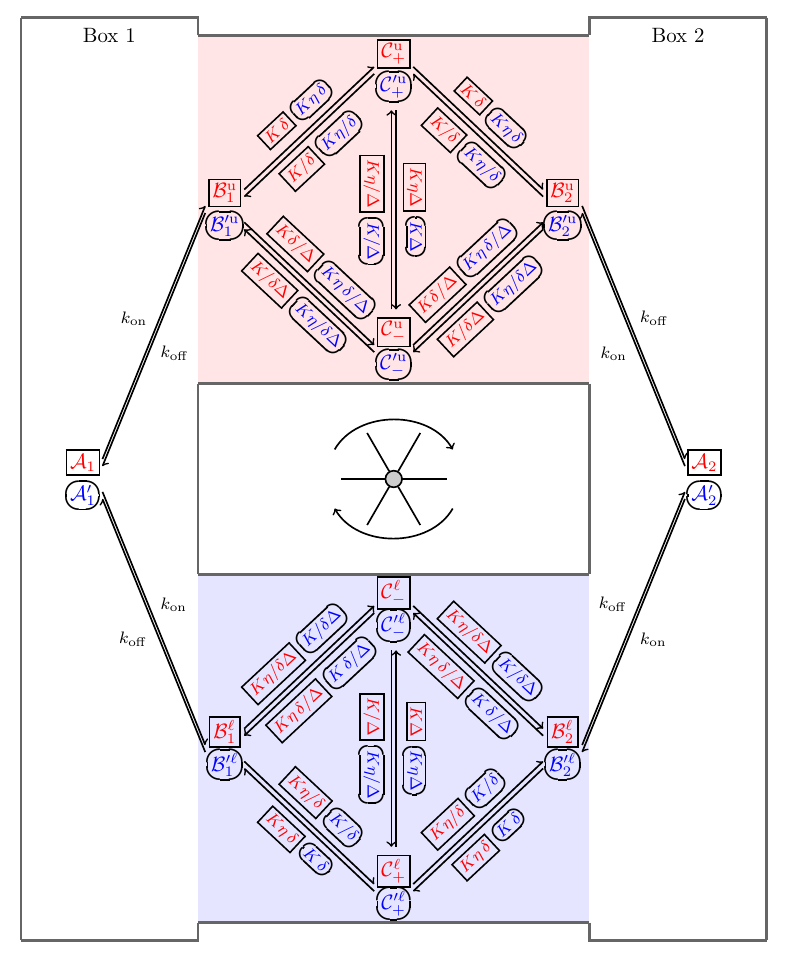}
	\caption{A modification of the kinetic network in Fig.~\ref{fig:KineticNet}, which employs additional branches in each channel. The rates in the ``$-$'' branches are suppressed by an active process which costs/produces free energy $\Delta F$ per transition. Notations are the same as in Fig.~\ref{fig:KineticNet}, but for compactness we denote $\delta\equiv e^{\Delta f}$ and $\Delta\equiv e^{\Delta F}$. {Note that, when $\Delta F=0$, a direct transition from one box to the other costs/produces free energy $2\Delta f$ -- twice as much work as the Fig.~\ref{fig:KineticNet} network. This choice was made to limit the profusion of factors of $\frac{1}{2}$, and it doesn't affect the spirit of the model.}}
	\label{fig:KineticNetActive}
\end{figure*}

A possible implementation of this is shown in Fig.~\ref{fig:KineticNetActive}. It extends the Fig.~\ref{fig:KineticNet} network with additional kinetic branches in each channel (there is now a ``$+$'' branch and a ``$-$'' branch), and also provides a route for transitioning between them. The rates in the ``$-$'' branches are reduced with respect to the ``$+$'' rates by an active process which consumes or produces free energy $\Delta F$ per transition.%
\footnote{It is worth emphasising here that there is no dissipation arising from the ``$-$'' branch alone. Dissipation instead arises because of the transition path between the two branches, which creates a loop with thermodynamic drive $2e^{\Delta F}$ -- the loop violates detailed balance and so the network must be driven.}
Furthermore, any particle in the ``$+$'' branch is more likely to be jump to the slower ``$-$'' branch than the other way around, meaning that the sorting process is slowed overall. Crucially, the rate of jumping between the branches of a given channel is different for different particle types -- this is the {discrimination} part. For a particle in the \textit{correct} channel (eg an unprimed particle in the upper channel), the rate of jumping between branches is small compared to the rate of transition between boxes, so particles in the correct channel will be relatively unperturbed by the extra active process. For a particle in the \textit{incorrect} channel, however, there will be a substantial flow from the fast ``$+$'' branch to the slow ``$-$'' branch. {Thus, particles are actively slowed when they are in the incorrect channel, which reduces the effective $\eta$ towards zero, and hence results in better sorting.}

For clarity, consider the limiting case where the ``incorrect'' particles are made so slow that they cannot make a transition between boxes on any reasonable time scale. Then the discrimination is nearly perfect, and particles may only travel towards their intended boxes.

\begin{figure}[ht]
	\centering
	\includegraphics[width=\wid\linewidth]{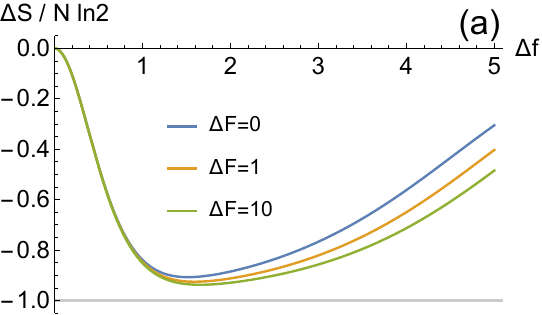}\\[1em]
	\includegraphics[width=\wid\linewidth]{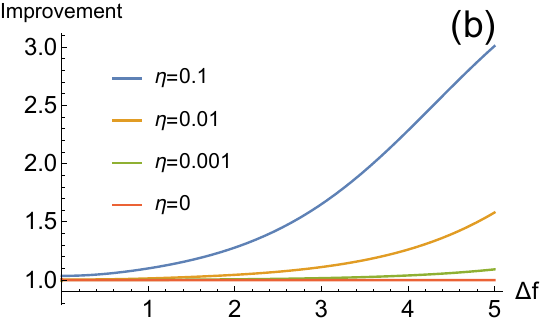}
	\caption{(a): Sorting quality $\Delta S$ plotted against driving $\Delta f$ for several values of the active discrimination $\Delta F$. The active discrimination improves the sort quality, especially when the driving $\Delta f$ is high. (b): The \textit{maximum} improvement engendered by active proofreading as a function of driving, for several error parameters $\eta$. This maximum improvement is defined as the ratio $\frac{\Delta S(\Delta F\to\infty)}{\Delta S(\Delta F=0)}$, and this quantity is seen to grow with $\eta$ as well as $\Delta f$.}
	\label{fig:ActiveKineticSortingEntropy}
\end{figure}

In Fig.~\ref{fig:ActiveKineticSortingEntropy}, we show that the modifications introduced in Fig.~\ref{fig:KineticNetActive} do indeed improve sorting above the $\Delta F=0$ baseline.

\begin{figure}[ht]
	\centering
	\includegraphics[width=\wid\linewidth]{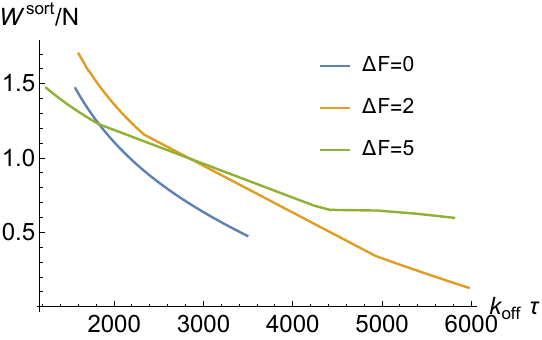}
	\caption{For a given purity of sorting (here $\Delta S=-0.4$), we plot the trade-offs between the work done to sort and the characteristic relaxation time of the network in Fig.~\ref{fig:KineticNetActive}. The error probability and the kinetic rates are fixed ($\eta=0.1$, $k_{\rm on}=10^{-3}k_{\rm off}$, and $K=0.1k_{\rm off}$), and the driving $\Delta f$ is varied to produce each line for a given $\Delta F$. (The kinks in the lines are an artifact from numerical calculation the system's relaxation time-scale.)}
	\label{fig:ActiveKineticTradeOff}
\end{figure}

An issue of relevance to real sorting devices, for instance the ones mentioned in the introduction, is the minimal work required to achieve a given entropy reduction, and the concomitant trade-offs with the time scales required to complete the sorting.%
\footnote{In regards to this last quantity, we shall consider the characteristic time-scales on which a given system evolves, rather than the full time taken to complete sorting (which may diverge).}
To resolve this matter for our system requires some numerical root-finding; but the example shown in Fig.~\ref{fig:ActiveKineticTradeOff} illustrates the fact that sorting can be completed quickly at the expense of additional work. Unexpectedly, it also demonstrates that strong active proofreading can in some cases \textit{reduce} the work needed to sort quickly.%
\footnote{When done right, speedy sorting decrements the leakage between boxes, and hence the rate of energy-consuming transitions.}

It is worth pointing out that, while we use our sorting device to sort a mixed system at the expense of work, it may equally well be run in reverse as an ``entropy engine'' which produces work from an initially segregated system. Two further regimes are accessible to this model: one where the network simultaneously sorts and ``produces'' work {(ie, consumes less work than the $\Delta F=0$ sorter)}, and a {``lose-lose'' regime} where the network consumes work to increase the entropy of the system.

\section{A Hopfield--Ninio Sorter Based on Energy Preference}
\label{sec:HNsorter}

\begin{figure}[ht]
	\centering
	\includegraphics[width=\wid\linewidth]{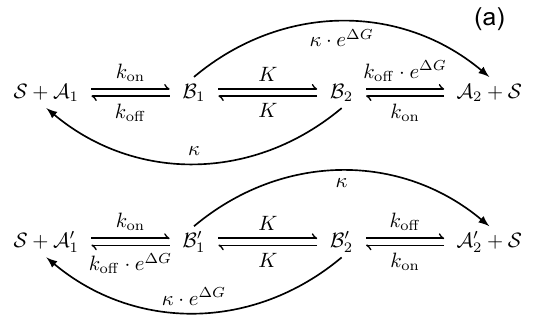}
	\includegraphics[width=\linewidth]{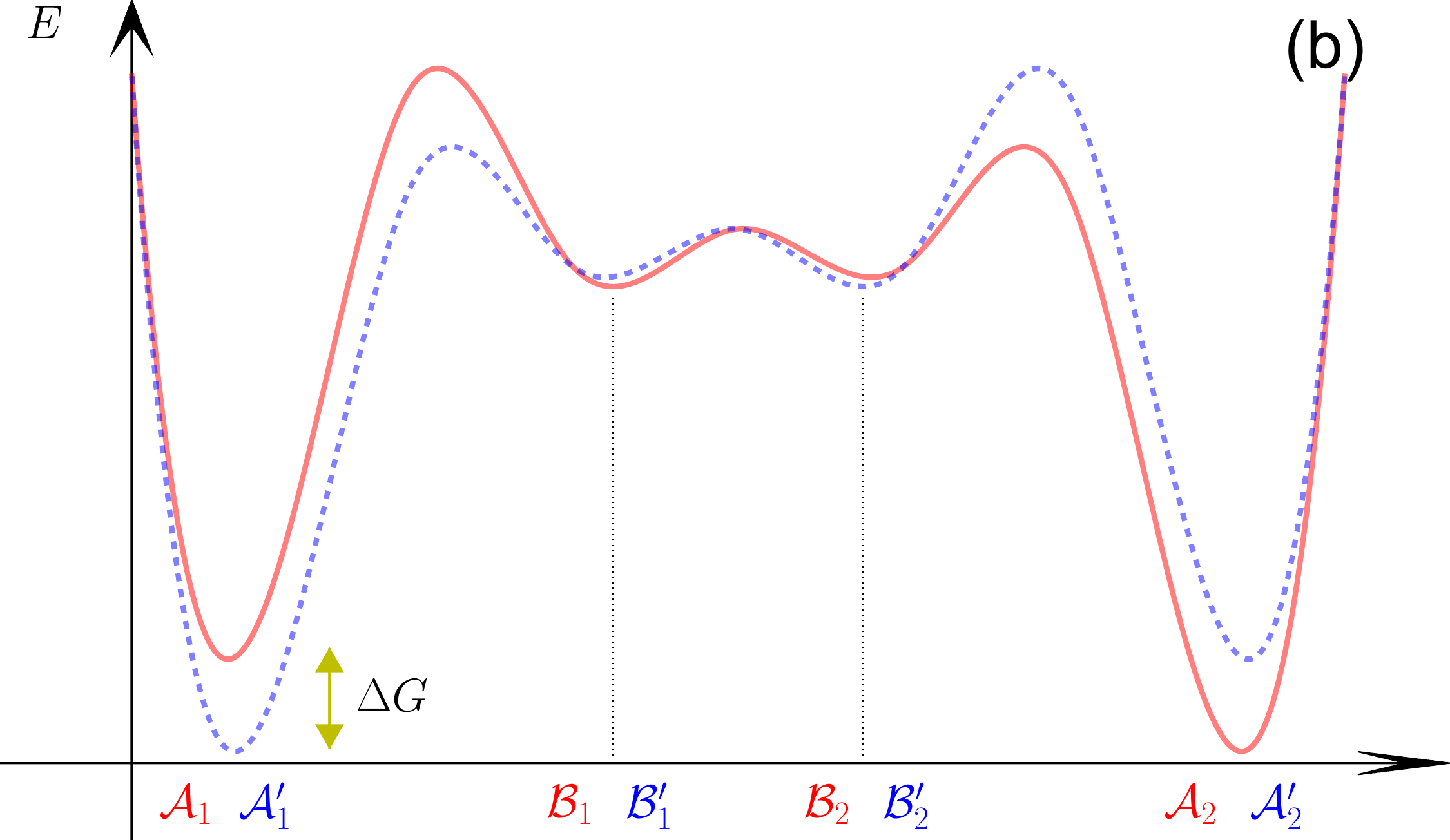}
	\caption{(a): The general reaction networks for the two species. The two networks share all rate constants except for those which differ by a factor of $e^{\Delta G}\geq 1$. One-way arrows consume free energy of $\Delta G$ per transition, and we assume that $\Delta G$ is so large that the rate of reverse transition is negligible on the time-scale of interest (see footnote~\ref{fnt:OneWayArrow}). Other choices for where to place the $e^{\Delta G}$ factors would work just as well for our purposes, but we follow the convention in \cite{Hop74}. (b): Energy landscape for the states in the network for both species, where $\mathcal A$ has the solid red curve and $\mathcal A^\prime$ has dashed blue. Note that, because we consider an active system, this landscape does not determine all the transition rates or the equilibrium state.}
	\label{fig:SymNet}
	\label{fig:EnergyLandscape}
\end{figure}

We now introduce our second model sorting device, which exploits \textit{energetic} differences between the particle types rather than just kinetic differences. Imagine that the particles are distinguished by their energetic preference for one box over another, and denote this energetic difference $\Delta G \geq 0$ (see Fig.~\ref{fig:EnergyLandscape}). {This $\Delta G$ therefore plays two overlapping roles: (i) it quantifies the distinguishability of the particles, and in this sense replaces $\eta$ from the previous model; and (ii) $\Delta G$ controls the equilibrium distribution of particles between the two boxes via the Boltzmann factor $e^{\Delta G}$. Thus, and in contrast to the previous model, the device already performs some sorting even in equilibrium.}%
\footnote{{It is tricky, therefore, to directly compare the two models using the same evaluation criteria. And, while this new model is interesting and practically relevant for sorting processes, it is a little further from the conventional Gibbs Paradox setup than the previous one (but we can still analyse how sorting quality varies with distinguishability -- see Fig.~\ref{fig:SSS}).}}
Here we seek to improve the quality of sorting by introducing active / dissipative processes into the system.%

We may now make a connection with a well-established body of literature. In the early seventies it was known that the accuracy of certain biological processes for distinguishing very similar particles (for example different nucleotides during RNA transcription) far exceeds the equilibrium expectation based on the enzyme-substrate binding energies. In response, Hopfield and Ninio independently developed dissipative ``proofreading'' schemes capable of drastically amplifying the existing binding energy difference \cite{Hop74,Nin75,MHL12}. It is natural for us to apply the Hopfield--Ninio proofreading scheme to our particle sorting problem, when there exists some difference $\Delta G$ in the free energy landscape of the two particle types.%

\begin{figure}[ht]
	\centering
	\includegraphics[width=0.7\linewidth]{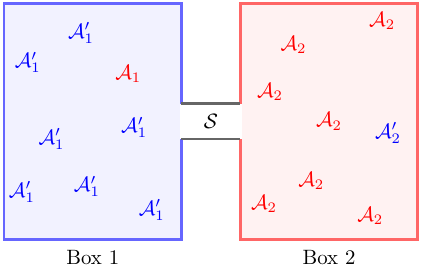}
	\caption{Sketch of the Hopfield--Ninio sorting device, in an imperfectly sorted steady state. The apparatus consists of two boxes connected by a single bidirectional channel and containing two types of particles, $\mathcal A$ and $\mathcal A^\prime$. When $\Delta G>0$, $\mathcal A$ accumulates in box~2, and $\mathcal A^\prime$ in box~1. See also the energy landscape sketched in Fig.~\ref{fig:EnergyLandscape}.}
	\label{fig:SymBox}
\end{figure}

Our model apparatus is sketched in Fig.~\ref{fig:SymBox}: a sorting device $\mathcal S$ sits in the single channel connecting two volumes, and may grant or deny passage through the channel depending on the outcome of a sorting process.

This process is modelled as the network sketched in Fig.~\ref{fig:SymNet}, which has four independent parameters: $\Delta G$ and four rate constants, one of which may be used to set the time-scale (but for clarity we leave all the rate constants explicit and with units).
When $\Delta G>0$, the sorting device promotes the accumulation of $\mathcal A$ in box~2 and $\mathcal A^\prime$ in box~1, as in section~\ref{sec:KineticSorting}. Every transition along one of the unidirectional edges dissipates $\Delta G$.%
\footnote{\label{fnt:OneWayArrow}These one-way arrows may look disturbing in light of our usual chemical kinetics experience, but remember, we are dealing here with a highly non-equilibrium, strongly driven system in which such processes are commonplace. See for instance the books \cite{Phillips_book,Bialek_book}, and the extensive literature on kinetic proofreading.}
Thus, when the kinetic rate $\kappa$ is zero, the network is an equilibrium sorter whose performance is ultimately controlled by the Boltzmann factor $e^{\Delta G}$.

\subsection{Steady State}
\label{sec:HNSS}

The system of nonlinear ODEs which represent this reaction network is written in Appendix~\ref{app:SSS}. As in the previous model, the steady state is exactly calculable (see Appendix~\ref{app:SSS}) and we can compute the entropy change for a given set of parameters as before. Better-than-equilibrium sorting can be achieved for any $\Delta G$, provided the kinetic rate of the energy-consuming transition, $\kappa$, is substantially smaller than the other rates, and $k_{\rm off}\gg k_{\rm on}$. The improvement is particularly impressive when the particles are highly distinguishable (ie $\Delta G$ is large), such that the Hopfield--Ninio sorter behaves like an equilibrium sorter with $\Delta G\to2\Delta G$.

We may also choose a different set of kinetic rates that cause us to pay work for sorting \textit{worse} than Boltzmann. This is possible when the rate $\kappa$ is large, and the sorting network encourages particles of both species to bypass the proofreading machinery in both directions via the energy-consuming transitions. These predictions for good and bad sorting regimes are shown to agree with simulations in Appendix~\ref{app:SSS}.

\subsection{Time and Work for Sorting}
\label{sec:HNTW}

It is plausible that the worse-than-equilibrium sorter just discussed sacrifices sorting quality in order to deliver improved speed, as has been found in other applications of Hopfield--Ninio-style networks \cite{MHL12,MHL14}. Our next task is to find the time-scale of the system's evolution, and compare it with the invested work. However, since the Hopfield--Ninio sorter model explicitly includes binding to the sorting device, the ODEs governing the evolution~(\ref{eqn:SymNetKinetics}) are non-linear, and to calculate anything beyond steady-state quantities requires some approximation.%
\footnote{The non-linearity is not an indispensible feature of an energy-driven sorting device such as the one we are consdering. But, although it does rather complicate calculations, we retain it in our model to make the connection with Hopfield--Ninio networks more obvious, and to enhance the relevance to practical applications such as cellular transport.}

Here we take $k_{\rm off}$ to be large compared to $K$ and $\kappa$: this allows us to treat the bound intermediate $\mathcal B$ states as evolving quasi-statically. This approach is standard in Michaelis--Menten kinetics, and allows us to compute the number of particles in the $\mathcal B$ states at every instant in terms of the number in the $\mathcal A$ states. Assuming the boxes are much larger than the channel which connects them, $\mathcal A_2 \approx N-\mathcal A_1$, where $N$ is again the total number of particles. Then $\mathcal A_1$ obeys the ODE
$\dot{\mathcal A_1} = \frac{1-\mathcal A_1/\mathcal A_1^{\rm SS}}{\gamma+\delta \mathcal A_1/N}$,
where $\mathcal A_1^{\rm SS}$ is the (exactly known) value of $\mathcal A_1$ in the steady state, and the positive constants $\gamma$ and $\delta$ have units of time. This can be solved to compute the evolution $t(\mathcal A_1)$ in terms of $\Delta G$ and the kinetic rates (see equation~(\ref{eqn:t(A)})).

Though this $t(\mathcal{A}_1)$ expression is not invertible, we may identify the dominant relaxation time as $\tau\equiv \mathcal A_1^{\rm SS}\left(\gamma+\delta\mathcal A_1^{\rm SS}/N\right)$ (this is extensive in the system size). One interesting limit of this $\tau$ is $\tau\sim 2/{k_{\rm off}e^{\Delta G}}$ when $\Delta G$ is large. This hints that there exist regimes where a better-than-equilibrium sorter can operate {faster} than a worse-than-equilibrium sorter; but only \textit{if} the particles are highly distinguishable (contrast with \cite{MHL12,MHL14} which found that high error rates are typically redeemed by enhanced speed).

A final quantity of interest is the total work required to sort the system into its steady state. One can obtain an expression for the {minimal%
\footnote{\label{fnt:MinimalDissipationRate} This is the \textit{minimal} rate if we assume each energy-consuming transition costs $\Delta G$ of free energy.}
}
rate of dissipation $\dot W$ as a function of the instantaneous $\mathcal A_1$, and integrate this over time to find the total work done (though only approximately -- see the discussion in Appendix~\ref{app:Dissipation}). While the full expression is a little unpleasant, it is much simplified in the regime when $\Delta G$ is very large, in which case ${W^{\rm sort}}/{N} \simeq \kappa\Delta G/k_{\rm off}$ (ie the work to sort increases linearly with the distinguishability). The full result is compared with simulations in Appendix~\ref{app:Dissipation}.

\subsection{Cost of a Desired Purity}
\label{sec:Phase}

\begin{figure}[ht]
	\centering
	\includegraphics[width=\wid\linewidth]{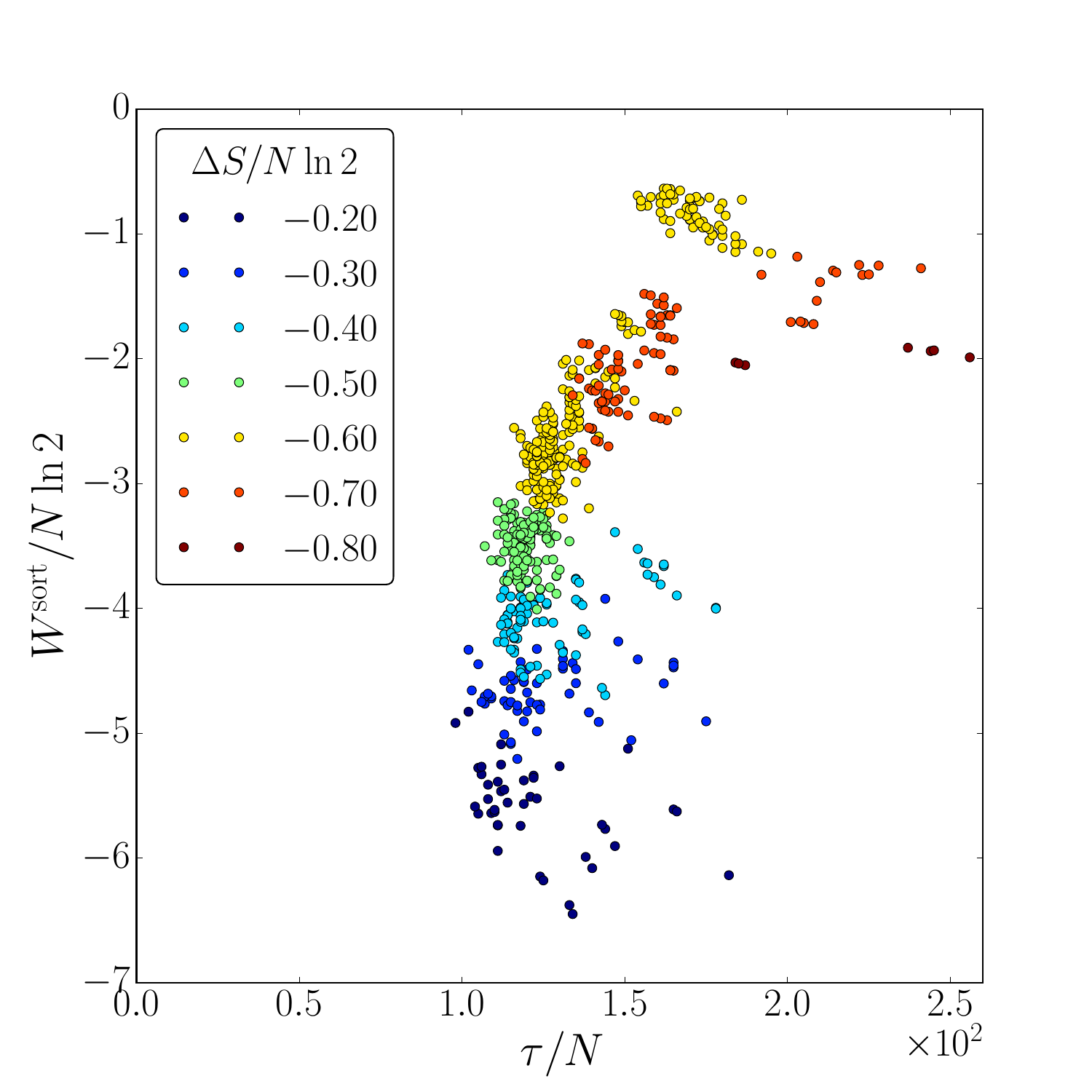}
	\caption{Entropy change from a simulation of the Fig.~\ref{fig:SymNet} network (in colour), plotted against the time and work to reach the steady state. Each point corresponds to a simulation of the network in Fig.~\ref{fig:SymNet} with different kinetic rates. $\Delta G=\ln10$ is fixed. The sorting quality tends to increase for smaller vaules of $W^{\rm sort}$.}
	\label{fig:Trade_D10}
\end{figure}

The foregoing findings allow us finally to find the costs to maintain a desired quality of sorting for a given particle distinguishability $\Delta G$. All the trade-offs we have met so far, between the quality of the sort, the time to reach the steady state, and the minimal work performed in sorting, can be projected onto a diagram such as Fig.~\ref{fig:Trade_D10}. This confirms that high-quality sorting (colours closer to red) needn't cost more work than low-quality sorting. The key message to take from this diagram, however, is that for a given sorting quality, high speeds cost more work. This finding is robust for other values of $\Delta G$.

\section{Conclusion}
\label{sec:conclusion}

The Gibbs Mixing Paradox is traditionally invoked to illustrate the role of particle distinguishability in the macrostates of statistical mechanics. In this paper, {we skirted the debate about the ``paradox'', which is resolved by acknowledging that distinguishability is either included in the system's macrostate or it is not. Instead, we struck a new path and considered a practical consequence of Gibbs' setup, which becomes apparent once we dispense with the idealisation of distinguishability: namely, we} investigated the performance of sorting devices which can \textit{partially} distinguish two similar types of particles. We then showed how, when such devices are allowed to dissipate energy, they may achieve sorting efficiencies surpassing (or otherwise) those of their passive, equilibrium counterparts.

Such processes are relevant to a variety of biological systems, whose function depends on maintaining a level of purity with respect to their environment -- we may consider organisms' ability to isolate and expel/metabolise contaminants, or the segregation of sodium and potassium ions across cellular membranes. Our aim here was to elucidate the physics which underlies such energy-consuming processes.

For concreteness, we introduced specific models based on two different sorting mechanisms, where the similarity of the particles to be sorted was represented by a single parameter. We showed how the efficacy of sorting was a continuous function of the particle similarity, and that it could be improved with the inclusion of active processes which effectively enhance the particle distinguishability. Furthermore, we found for both models that accurate sorting can be achieved quickly and with very low dissipation for carefully selected model parameters; but any improvement in speed must generally be paid for with more work, and vice-versa.

Considering the models of particle sorting, we demonstrated that Gibbs paradox, as a ``paradox'', arises solely from the unwarranted exploitation of some of the idealizations of equilibrium statistical mechanics.  The idealized statements are undoubtedly correct, such as ``if the particles are distinguishable \textit{in principle}, then a device \textit{in principle} can be built to sort particles at the expense of work $k_{\rm B}T \ln 2$ per particle (or that this work can be extracted \textit{in principle} by allowing particles to mix) -- there is no paradox here.  But if we start thinking how to sort particles in reality, then we realize that applicability of such concepts of statistical mechanics becomes increasingly stringent as particles become increasingly similar, and sorting them within a finite amount of time and by a practically realizable device may require greater work expenditure, which only approaches equilibrium statistical mechanics prescriptions in the right limit.  While this conclusion may be viewed trivial with hindsight, it in nevertheless a worthy exercise to demonstrate it on specific examples of practical relevance.

\begin{acknowledgements}
	This work was supported primarily by the MRSEC Program of the National Science Foundation under Award Number DMR-1420073.  AYG acknowledges stimulating discussions with R.~Phillips.
\end{acknowledgements}


\appendix
\def\wid{0.9}

\section{Entropy Definition}
\label{app:Entropy}

For both sorting models (sections~\ref{sec:KineticSorting} and~\ref{sec:HNsorter}), the quality of sorting is measured as an entropy change from the initial, maximally-disordered state to the partially-ordered steady state. Since we consider sorting devices which treat particle type $\mathcal A$ identically to type $\mathcal A^\prime$ (with box~1 and box~2 switched), we can restrict attention to particle type $\mathcal A$ alone.

The total entropy per particle is calculated from the number of $\mathcal A$ particles in box~1, $\mathcal A_1$, using the formula
\begin{align}
	\frac{\Delta S}{N} = -\frac{\mathcal A_1}{N}\ln\left(\frac{\mathcal A_1}{N}\right) - \frac{N-\mathcal A_1}{N}\ln\left(\frac{N-\mathcal A_1}{N}\right) - \ln2 \ ,
		\label{eqn:Entropy}
\end{align}
where $N$ is the total number of particles, and we assume the capacity of the boxes is much larger than that of the sorting device which connects them (so $\mathcal A_2\approx N-\mathcal A_1$).

With this definition, the entropy change for no sorting is zero, while perfect sorting would change the entropy by $-\ln2$ per particle.

\section{Hopfield--Ninio Sorter Steady State}
\label{app:SSS}

The kinetics for the $\mathcal A$-type particles in the Fig.~\ref{fig:SymNet} is given by:
\begin{align}
	\begin{split}
		\dot{\mathcal A_1} &= -k_{\rm on}\cdot \mathcal S\cdot \mathcal A_1 + k_{\rm off}\cdot \mathcal B_1 + \kappa\cdot \mathcal B_2 \\
		\dot{\mathcal B_1}   &= +k_{\rm on}\cdot \mathcal S\cdot \mathcal A_1 - (k_{\rm off}+K+\kappa\cdot e^{\Delta G})\cdot \mathcal B_1 + K\cdot \mathcal B_2 \\
		\dot{\mathcal B_2}   &= +K  \cdot \mathcal B_1 - (k_{\rm off}\cdot e^{\Delta G}+K+\kappa)\cdot \mathcal B_2 + k_{\rm on} \cdot \mathcal S\cdot \mathcal A_2 \\
		\dot{\mathcal A_2} &= +\kappa\cdot e^{\Delta G}\cdot \mathcal B_1 + k_{\rm off}\cdot e^{\Delta G}\cdot \mathcal B_2 - k_{\rm on}\cdot \mathcal S\cdot \mathcal A_2 \ ,
	\end{split}
	\label{eqn:SymNetKinetics}
\end{align}
where $\mathcal A_1$ denotes the number of particles in box 1, etc. A similar system of equations obtains for the $\mathcal A^\prime$ particles.

There is an additional constraint, inherited from the Hopfield--Ninio proofreading scheme, that $\mathcal S$ can only sort a finite number of particles at once. Calling the maximum $\mathcal S_0$, we have
\begin{align}
	\mathcal S = \mathcal S_0 - \mathcal B_1 - \mathcal B_2 - \mathcal B_1^\prime - \mathcal B_2^\prime \ .
	\label{eqn:SConstraint}
\end{align}
The presence of $\mathcal S$ in equation~(\ref{eqn:SymNetKinetics}) makes the system nonlinear, and also couples the dynamics of the two networks. However, the symmetry between the two types of particles allows us to mostly avoid the complications of coupling.

The steady state occupation of either box can be easily computed by setting all the time derivatives in equations~(\ref{eqn:SymNetKinetics}) to zero, and solving algebraically. The result is
\begin{align}
	\frac{\mathcal A_1^{\rm SS}}{N} = \frac{c_1+\kappa^2e^{\Delta G}}{c_1+c_2\,e^{\Delta G}+k_{\rm off}\kappa e^{2\Delta G}} \ ,
	\label{eqn:ASS}
\end{align}
where $c_1\equiv k_{\rm off}\kappa+k_{\rm off}K+\kappa K$ and $c_2 \equiv 2\kappa^2+k_{\rm off}K+\kappa K$.

Due to the symmetry we imposed on our model, equation~\ref{eqn:ASS} accounts for other steady state densities via $\mathcal A_1=\mathcal A^\prime_2=N-\mathcal A_2=N-\mathcal A^\prime_1$. The network discussed here promotes the state $\mathcal A_2$ over $\mathcal A_1$, so good sorting will result in $\mathcal A_1$ close to zero. Note that for $\Delta G=0$, $\mathcal A_1^{\rm SS}={N/2}$ as expected.

For an equilibrium sorter (with $\kappa=0$), equation~(\ref{eqn:ASS}) becomes ${\mathcal A_1^{\rm eq}}/{N}=\left(1+e^{\Delta G}\right)^{-1}$. This is the correct Boltzmann result, and is naturally independent of kinetic coefficients. As noted in section~\ref{sec:HNSS} of the main text, equation~(\ref{eqn:ASS}) tells us that the quality of the active sorter may be much better than an equilibrium sorter when $\kappa$ is substantially smaller than the other rates, and $k_{\rm off}$ is simultaneously large enough to support the $e^{2\Delta G}$ term in the denominator.

\begin{figure}[ht]
	\centering
	\includegraphics[width=\wid\linewidth]{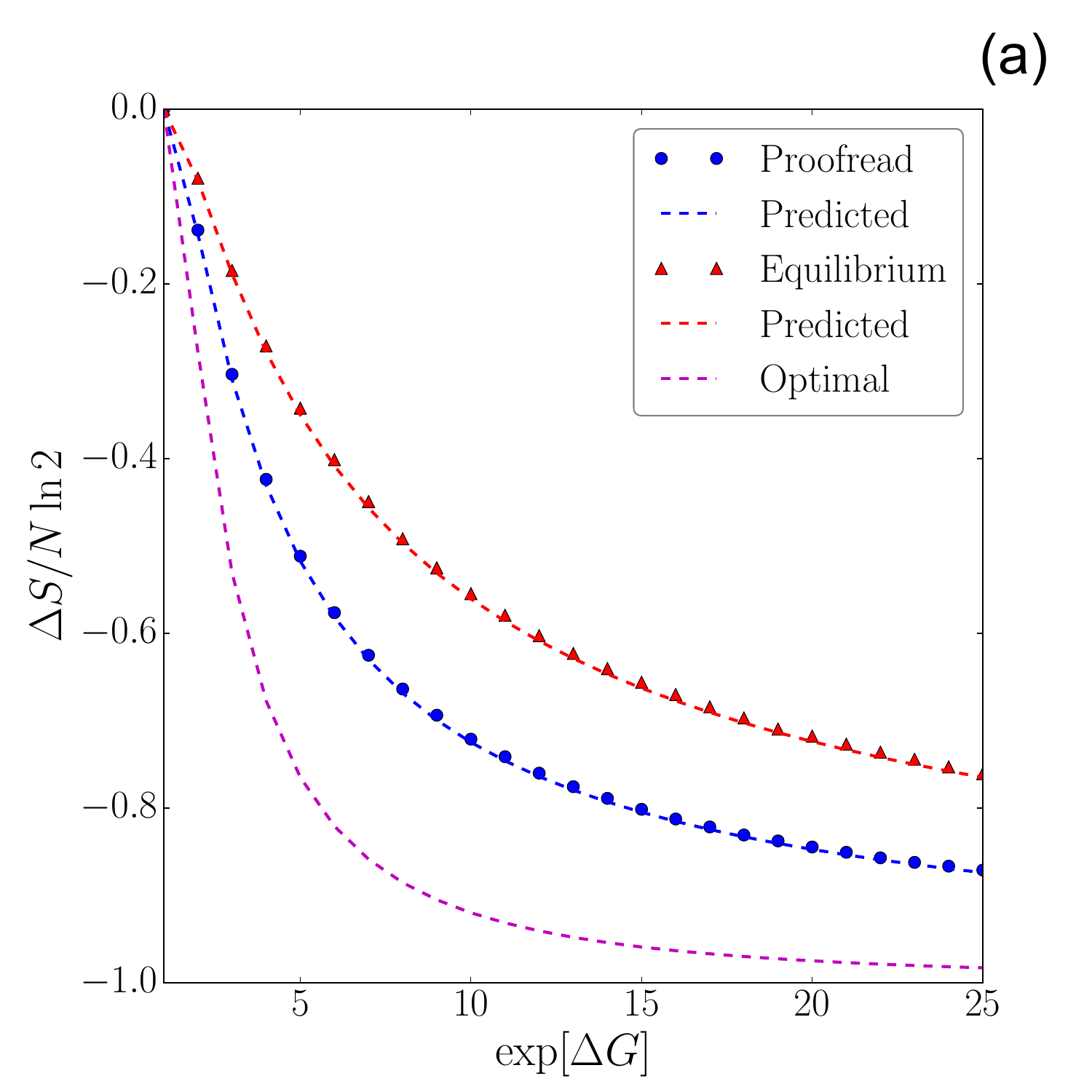}
	\includegraphics[width=\wid\linewidth]{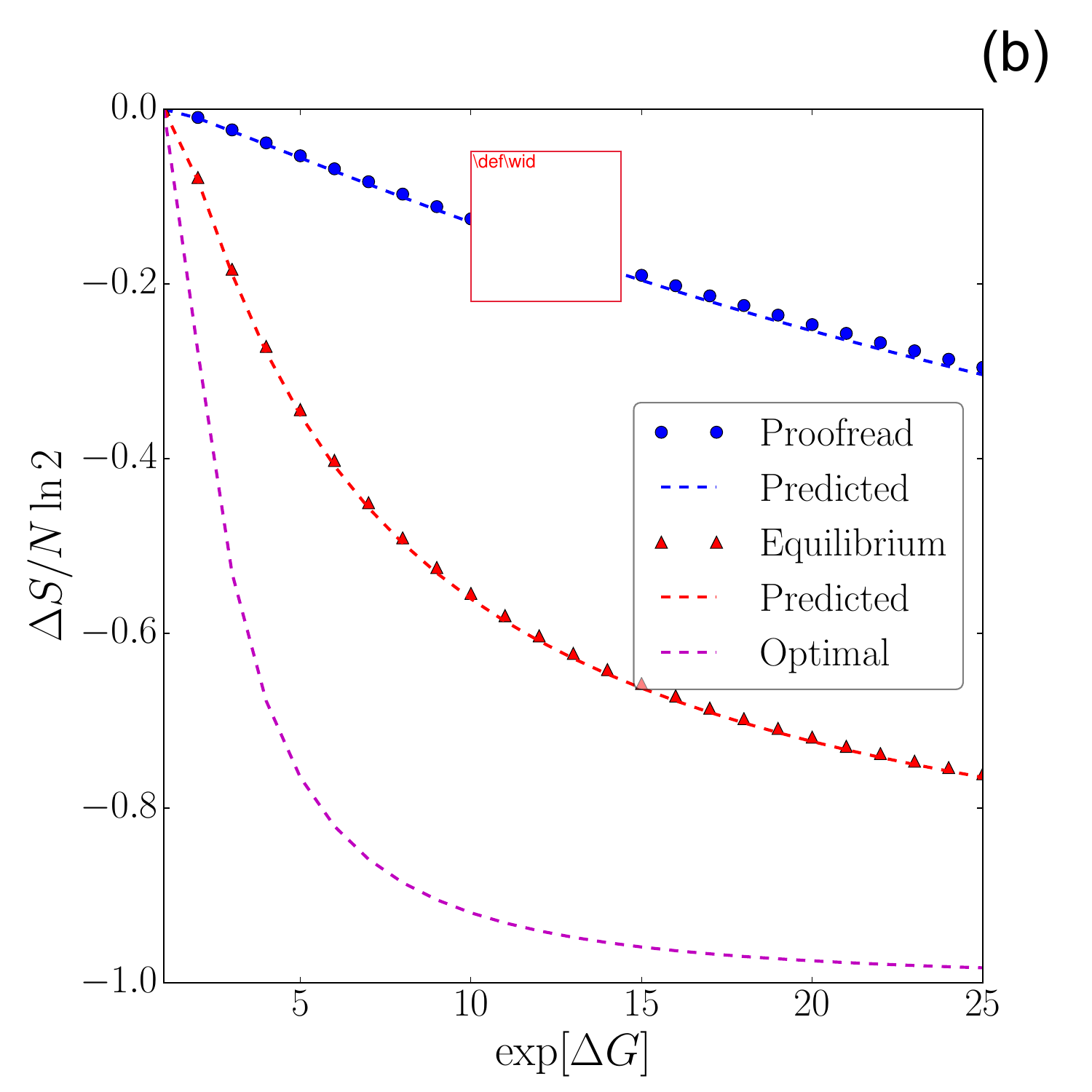}
	\caption{The steady state entropy $S$ as a function of the distinguishability parameter $\Delta G$, compared with the theoretical prediction from equation~(\ref{eqn:ASS}), and the Boltzmann value. The entropy is quoted in units of $\ln2$, and low values correspond to accurate sorting. (a): a better-than-equilibrium sorter; (b): a worse-than-equilibrium sorter.}
	\label{fig:SSS}
\end{figure}

From equation~(\ref{eqn:ASS}), we also observe that the active sorter performs \textit{worse} than a Boltzmann sorter for certain parameter choices. For instance, if we consider high $\Delta G$, then when $\kappa$ is of order $K$, but $k_{\rm off}$ is much smaller than $K/e^{\Delta G}$, we have ${\mathcal A_1^{\rm SS}}/{N} \sim {1}/{3}$, which is larger than the equilibrium value for $\Delta G>\ln2$. This is also illustrated in Fig.~\ref{fig:SSS}. In this case the network encourages particles of both species to bypass the proofreading machinery via the energy-consuming transitions, such that we pay work for worse sorting. In Appendix~\ref{app:tSS} we find that the time to perform the sorting \textit{may} be reduced by sacrificing sort quality (since particles avoid rattling around in the heart of the network). However, this sacrifice is not absolutely necessary, and accurate sorting can be achieved quickly if parameters are chosen judiciously.

To verify equation~(\ref{eqn:ASS}) and the results of the following sections, we perform stochastic simulations of the Fig.~\ref{fig:SymNet} network (the details are described in Appendix~\ref{app:HNSimulation}). Figure~\ref{fig:SSS} shows the steady state entropy of the system as a function of the distinguishability parameter $\Delta G$ for two choices of kinetic rates -- corresponding to a better-than-equilibrium active sorter and a worse-than-equilibrium active sorter. We find good agreement with our prediction, and see clearly how the discontinuity of the traditional Gibbs Mixing Paradox is softened, with $S$ approaching $-1$ asymptotically as $\Delta G\to\infty$.

\section{Hopfield--Ninio Sorting Time}
\label{app:tSS}

In section~\ref{sec:HNTW}, we describe the ``intermediate steady state'' approximation for calculating the non-linear system's dynamics. This, along with the ``large box'' assumption described in Appendix~\ref{app:Entropy}, yields the un-coupled ODE for $\mathcal A_1$:
\begin{align}
	\dot{\mathcal A_1} = \frac{1-\mathcal A_1/\mathcal A_1^{\rm SS}}{\gamma+\delta \mathcal A_1/N} \ ,
	\label{eqn:AdotIntSS}
\end{align}
where $A_1^{\rm SS}$ is given by equation~(\ref{eqn:ASS}), and the known constants $\gamma$ and $\delta$ are positive, intensive and have units of time. Equation~(\ref{eqn:AdotIntSS}) can be solved for the evolution:
\begin{align}
	t(\mathcal A_1) = \theta\cdot\left(\frac{1}{2}-\frac{\mathcal A_1}{N}\right) + \tau\cdot\ln\left(\frac{{N}/{2\mathcal A_1^{\rm SS}}-1}{{\mathcal A_1}/{\mathcal A_1^{\rm SS}}-1}\right) \ ,
	\label{eqn:t(A)}
\end{align}
where we've used the maximum-entropy initial condition $\mathcal A_1(t=0)=\frac{N}{2}$. Thus there are two effective time-scales, $\theta \equiv \delta A_1^{\rm SS}$ and $\tau\equiv \mathcal A_1^{\rm SS}\left(\gamma+\delta\mathcal A_1^{\rm SS}/N\right)$ (using the constants introduced in equation~(\ref{eqn:AdotIntSS})). Both $\theta$ and $\tau$ are extensive in the system size $N$. Unfortunately, $t(\mathcal A_1)$ is not invertible.

The total time to reach the steady state \textit{should} correspond to evaluating equation~(\ref{eqn:t(A)}) at $\mathcal A_1=\mathcal A_1^{\rm SS}$; however, the second term diverges to positive infinity at this point, reminding us that the steady state is reached only asymptotically (as one might intuitively expect when we invoke $N\gg1$ in oreder to treat particle number as continuous).
Because of the divergence of the term associated with $\tau$, it makes sense to provisionally identify $\tau$ as the dominant time-scale in the problem.%
\footnote{Note that choosing $\tau$ as the time-scale associated with the dominant term in equation~(\ref{eqn:t(A)}) does not necessarily mean that $\tau$ is greater than $\theta$.}

While the full expression for $\tau$ in terms of kinetic coefficients and $\Delta G$ is long and not particularly illuminating, some features are easy to understand. We already noted in the main text that $\tau(\Delta G\gg 0)\sim \frac{N}{S_0}\frac{2}{k_{\rm off}e^{\Delta G}}$. In general when $N$ is large, $\tau\sim \frac{N}{S_0}$ for any choice of parameters
Another interesting case is $\kappa=0$, which brings us back to a passive Boltzmann sorter. Then the time-scale is
\begin{align}
	\tau^{\rm eq}\simeq\frac{N}{S_0}\frac{4(K+(K+k_{\rm off})e^{\Delta G})}{k_{\rm off}K(1+e^{\Delta G})^2}.
\end{align}

\begin{figure}[ht]
	\centering
	\includegraphics[width=\wid\linewidth]{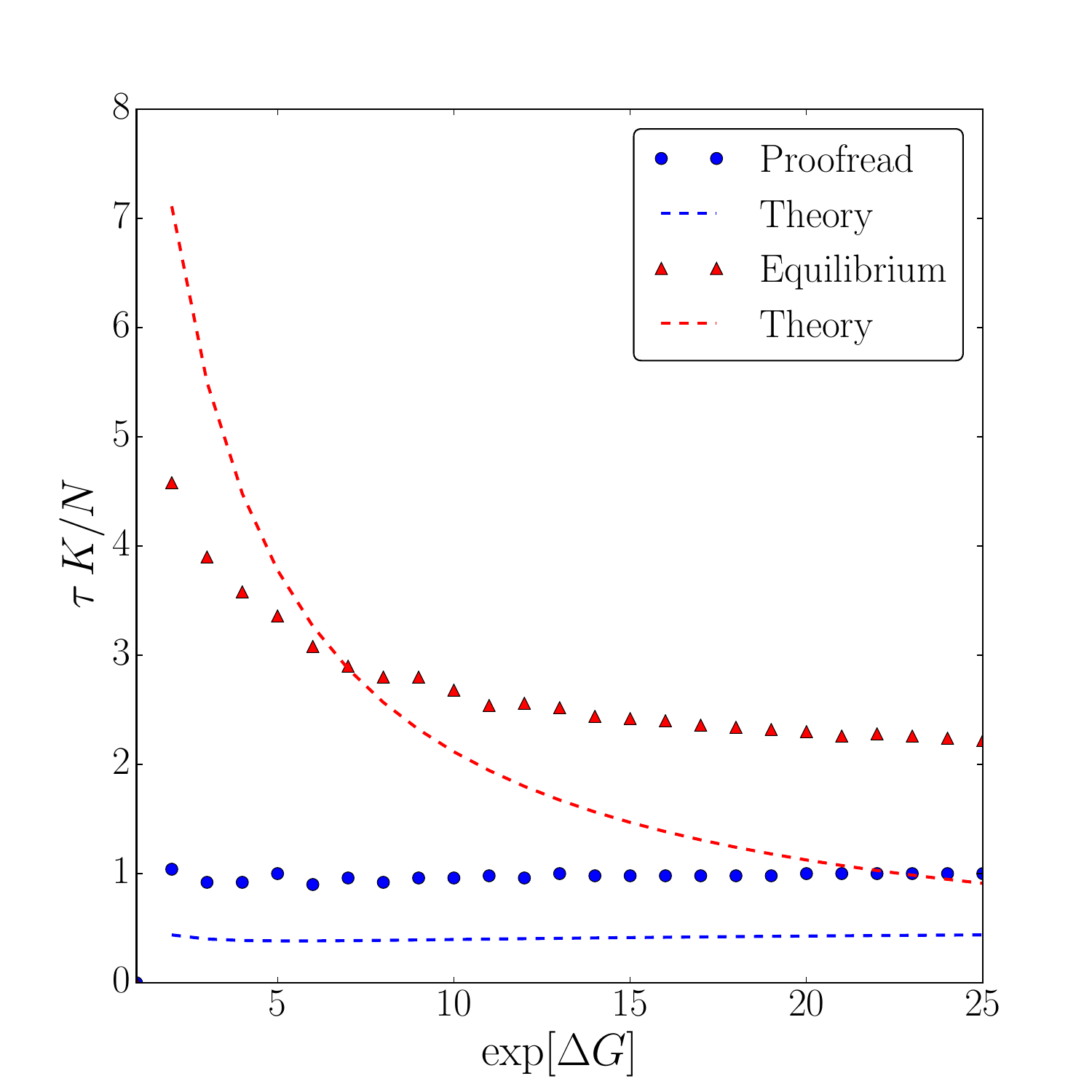}
	\caption{The time to reach the steady state as a function of $\Delta F$ for some choice of kinetic coefficients. The theoretical prediction for the time-scale $\tau$ is shown as dashed lines.}
	\label{fig:tSS}
\end{figure}

In Fig.~\ref{fig:tSS}, the full prediction for $\tau$ is plotted alongside the time to reach the steady state measured in simulations (see Appendix~\ref{app:HNSimulation}).

\section{Hopfield--Ninio Sorter Dissipation}
\label{app:Dissipation}

As stated in section~\ref{sec:HNTW}, we use the intermediate steady state approximation to find the {minimal%
\footnote{See footnote~\ref{fnt:MinimalDissipationRate}.}
} rate of dissipation $\dot W$ as a function of $\mathcal A_1$:
\begin{align}
	\dot W(\mathcal A_1) = \frac{\epsilon+\zeta\frac{\mathcal A_1}{\mathcal A_1^{\rm SS}}}{\gamma + \delta\frac{\mathcal A_1}{N}} \ ,
	\label{eqn:Wdot}
\end{align}
where $\epsilon, \zeta$ are known constants, and $\gamma, \delta$ are our friends from Appendix~\ref{app:tSS}. Dividing equation~(\ref{eqn:Wdot}) by equation~(\ref{eqn:AdotIntSS}), we get $\upd W/\upd \mathcal A_1$. In principle, integrating this from $\mathcal A_1=N/2$ to $\mathcal A_1^{\rm SS}$ gives the total work done to complete the sorting. However, the integral diverges, reflecting the fact that the steady state is only reached in the $t\to\infty$ limit (see Appendix~\ref{app:tSS}), and meanwhile work is constantly being done pushing particles in futile loops.
The work done to achieve $\mathcal A_1$ particles in box 1 is
\begin{align}\begin{split}
	W(\mathcal A_1) \approx \mathcal A_1^{\rm SS}&\left( \zeta\left(\frac{1}{2}-\frac{\mathcal A_1}{N} \right)+\right.\\
	&\;\left.+ \left( \epsilon + \zeta\frac{\mathcal A_1^{\rm SS}}{N} \right)\ln\left(\frac{\frac{N}{2\mathcal A_1^{\rm SS}}-1}{\frac{\mathcal A_1}{\mathcal A_1^{\rm SS}}-1}\right) \right) \ ,
		\label{eqn:Wsort}
\end{split}\end{align}
provided $\mathcal A_1^{\rm SS}\ll \mathcal A_1 \leq N/2$. We may evaluate this at say $\mathcal A_1=2\mathcal A_1^{\rm SS}$ to get an idea of the amount of work done to sort. The full expression is not terribly interesting, but, as noted in the main text, the high-$\Delta G$ regime yields a minimal work $W^{\rm sort}/N\approx \frac{\kappa}{k_{\rm off}} \Delta G$, which is zero when $\kappa=0$ (as it should be). It is perhaps surprising that more distinguishable particles require more work to sort; but this is simply because the dissipation of the non-equilibrium steps is commensurately greater.

\begin{figure}[ht]
	\centering
	\includegraphics[width=\wid\linewidth]{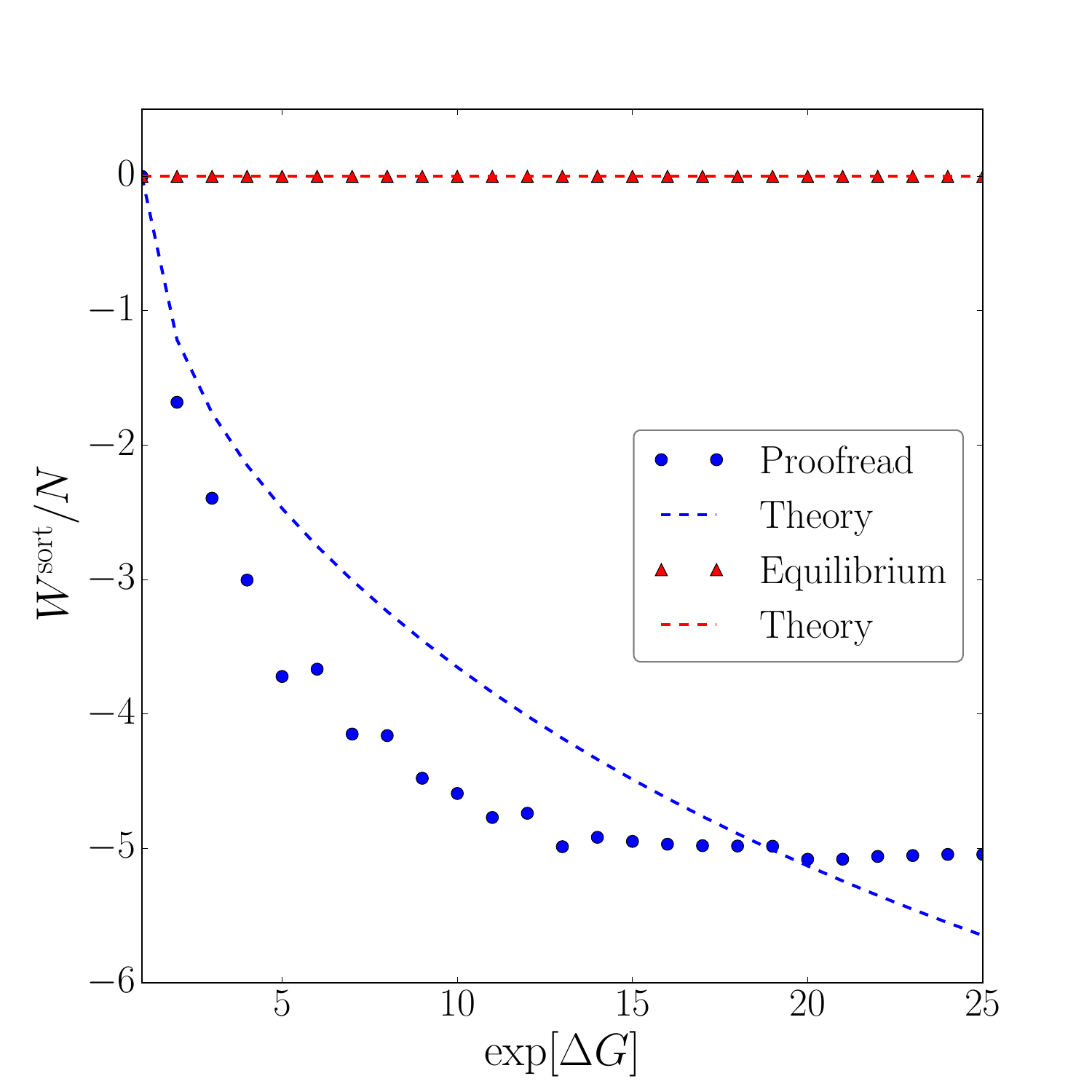}
	\caption{The total work required to reach the steady state: comparison of simulation and the equation~(\ref{eqn:Wsort}) approximation. Kinetic coefficients are the same as for Fig.~\ref{fig:tSS}.}
	\label{fig:Wsort}
\end{figure}

The full approximate calculation is shown alongside simulation data in Fig.~\ref{fig:Wsort}.

A potentially unwelcome feature of our sorting device is that is continues to consume energy even after the steady state has been reached. Using the full steady state densities in equation~(\ref{eqn:ASS}), we obtain the minimal rate of dissipation
\begin{align}
	\frac{\dot W^{\rm SS}}{N \Delta G} = \frac{\text{const}+k_{\rm on}\kappa\,e^{\Delta G}}{\text{const} +\text{const}\cdot e^{\Delta G} + k_{\rm off} e^{2\Delta G}} \ ,
	\label{eqn:WdotSS}
\end{align}
where the constants, which have been omitted for compactness, depend on the kinetic rates. If we examine the high-$\Delta G$ regime of equation~(\ref{eqn:WdotSS}), we find again that for a ``good sorter'' (with $k_{\rm off}$ large compared to $\kappa$), the dissipation in the steady state is suppressed and falls more quickly with $\Delta G$.

\begin{figure}[ht]
	\centering
	\includegraphics[width=\wid\linewidth]{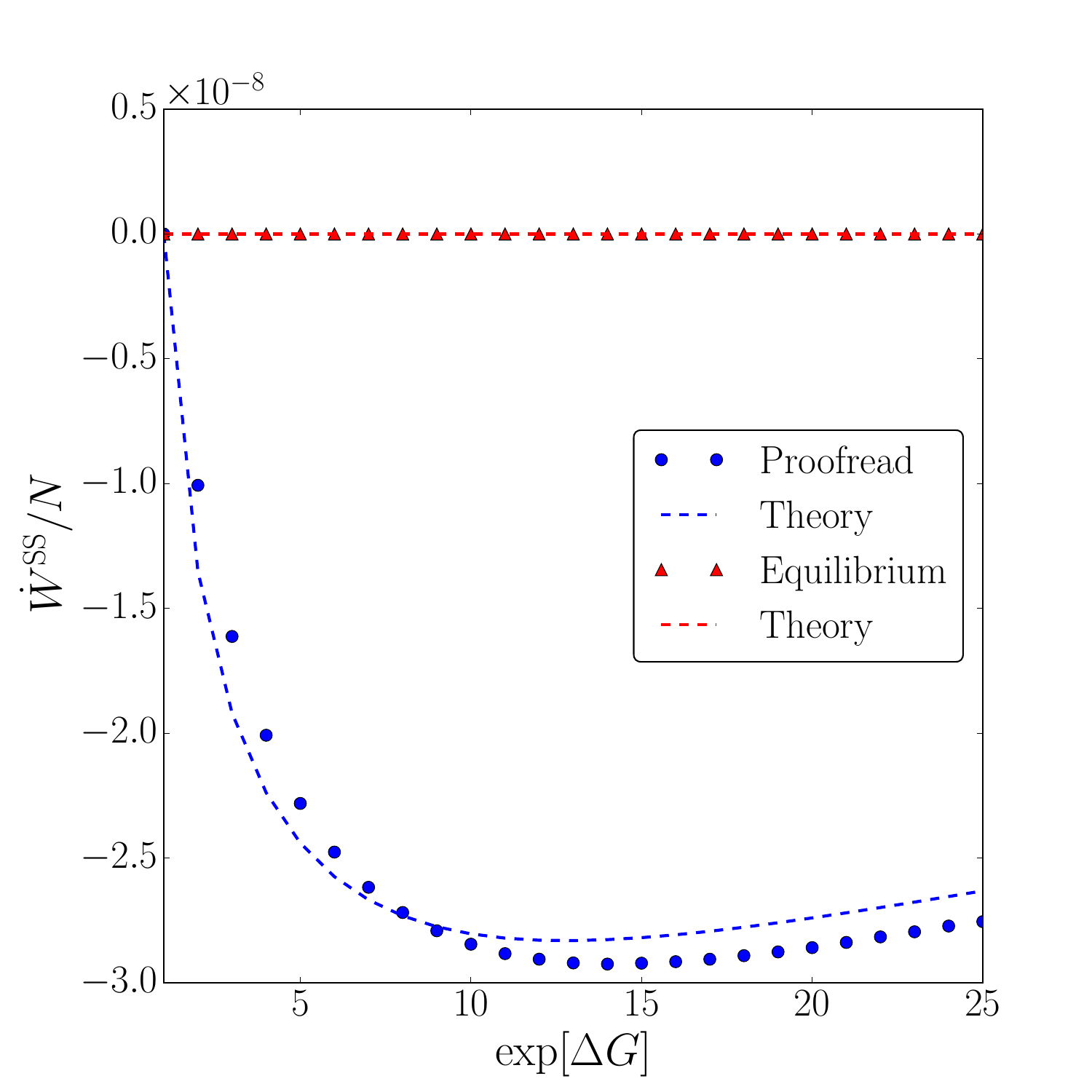}
	\caption{Dissipation rate in the steady state: comparison of simulation and the equation~(\ref{eqn:Wdot}) approximation. Kinetic coefficients are the same as for Fig.~\ref{fig:tSS}.}
	\label{fig:WdotSS}
\end{figure}

The simulation data matches the predicted trend, as seen in Fig.~\ref{fig:WdotSS}. Also visible in both plots is the competition between accurate discrimination reducing number of unnecessary dissipative transitions, and the energy cost of each transition: for smaller $\Delta G$ the latter (linear) dominates, while at higher $\Delta G$ the (exponential) discrimination wins and reduces the work cost.

\section{Hopfield--Ninio Sorter Simulations}
\label{app:HNSimulation}

The network in Fig.~\ref{fig:SymNet} was simulated using a ``time-triggered'' stochastic procedure: at each time step, the sorting device decides with some probability whether to ``bind'' to a new particle if it is unoccupied, or whether to progress an already-bound particle along the reaction chain.

\begin{figure}[ht]
	\centering
	\includegraphics[width=\wid\linewidth]{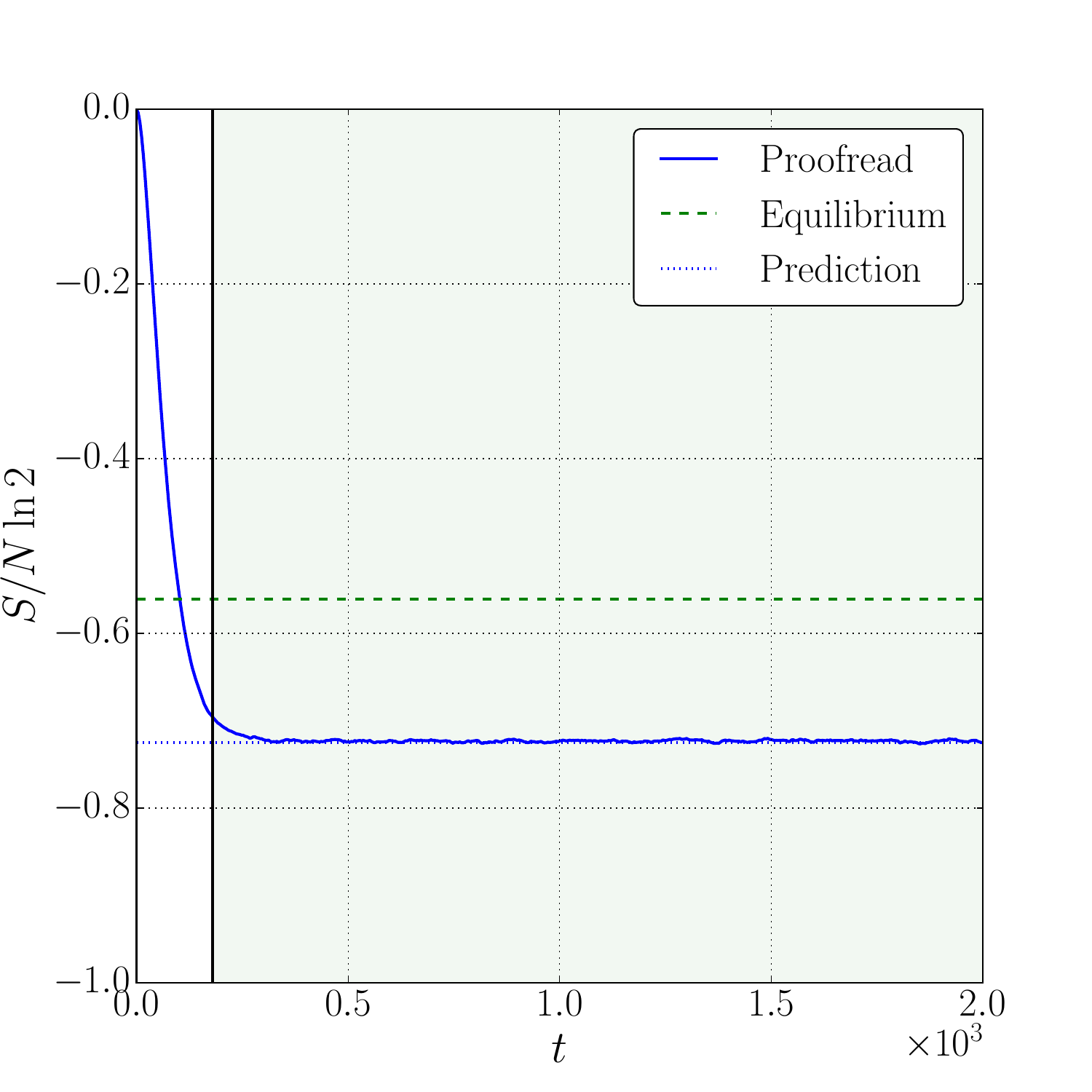}
	\caption{System entropy for the Hopfield--Ninio sorter as a function of time (blue solid line), which converges to and fluctuates around the prediction for steady state value (blue dotted). The free energy difference is $\Delta F=\ln10$, and the steady state entropy for an equilibrium sorter is shown as a green dashed line, while the onset of the steady state (found algorithmically) is marked with a vertical black solid line.}
	\label{fig:S(t)}
\end{figure}

Starting from the maximum-entropy initial condition, the simulation continues until the steady state is reached. See for example Fig.~\ref{fig:S(t)}, which shows the value of the entropy as a function of time for some choice of parameters.

\begin{thebibliography}{10}

\bibitem{Gibbs1}
J.~W. Gibbs.
\newblock On the equilibrium of heterogeneous substances.
\newblock {\em Trans. Connect. Acad. Sci.}, 3:108--248, 1876.

\bibitem{Gibbs2}
J.~W. Gibbs.
\newblock On the equilibrium of heterogeneous substances.
\newblock {\em Trans. Connect. Acad. Sci.}, 3:343--524, 1876.

\bibitem{Gib02}
J.~W. {Gibbs}.
\newblock {\em {Elementary Principles of Statistical Mechanics}}.
\newblock Dover, New York, 1960.

\bibitem{LandauLifshitz_StatisticalPhysics}
Lev~D. Landau and Evgenii~M. Lifshitz.
\newblock {\em Statistical Physics, Part 1 (Course of Theoretical Physics,
  Volume 5)}.
\newblock Butterworth-Heinemann, 3rd edition edition, 1980.

\bibitem{Hop74}
J.~J. Hopfield.
\newblock Kinetic proofreading: A new mechanism for reducing errors in
  biosynthetic processes requiring high specificity.
\newblock {\em Proceedings of the National Academy of Sciences},
  71(10):4135--4139, 1974.

\bibitem{Nin75}
J.~Ninio.
\newblock Kinetic amplification of enzyme discrimination.
\newblock {\em Biochimie}, 57(5):587 -- 595, 1975.

\bibitem{W+N03}
Matthias Weiss and Tommy Nilsson.
\newblock A kinetic proof-reading mechanism for protein sorting.
\newblock {\em Traffic}, 4(2):65--73, 2003.

\bibitem{Sa+06}
I.~{Sachs}, S.~{Sen}, and J.~{Sexton}.
\newblock {\em Elements of statistical mechanics with an introduction to
  quantum field theory and numerical simulation}.
\newblock Cambridge University Press, Cambridge, UK New York, 2006.

\bibitem{Mat08}
D.~C. {Mattis}.
\newblock {\em Statistical mechanics made simple.}
\newblock World Scientific, New Jersey, 2nd ed. / daniel c. mattis, robert h.
  swendsen. edition, 2008.

\bibitem{Kar07}
M.~Kardar.
\newblock {\em Statistical physics of particles}.
\newblock Cambridge University Press, Cambridge, 2007.

\bibitem{For13}
I.~Ford.
\newblock {\em Statistical physics an entropic approach}.
\newblock Wiley, Chichester, 2013.

\bibitem{Da+11}
N.~Dalarsson, M.~Dalarsson, and L.~Golubović.
\newblock {\em Introductory statistical thermodynamics}.
\newblock Academic Press, Amsterdam, 2011.

\bibitem{Stu03}
M.~D. Sturge.
\newblock {\em Statistical and thermal physics : fundamentals and
  applications}.
\newblock A.K. Peters, Natick, Mass., 2003.

\bibitem{Tuc10}
M.~E. Tuckerman.
\newblock {\em Statistical mechanics theory and molecular simulation}.
\newblock Oxford University Press, Oxford, 2010.

\bibitem{Hua87}
K.~Huang.
\newblock {\em Statistical mechanics}.
\newblock Wiley, New York, 2nd ed. edition, 1987.

\bibitem{Rei98}
L.~E. Reichl.
\newblock {\em A modern course in statistical physics}.
\newblock John Wiley, New York, second edition edition, 1998.

\bibitem{Pau73}
W.~Pauli.
\newblock {\em Thermodynamics and the kinetic theory of gases.}
\newblock MIT Press, Cambridge, Mass., 1973.

\bibitem{B-N07}
A.~{Ben-Naim}.
\newblock {On the So-Called Gibbs Paradox, and on the Real Paradox}.
\newblock {\em Entropy}, 9:132--136, September 2007.

\bibitem{Swe15}
R.~H. {Swendsen}.
\newblock {The ambiguity of ''distinguishability'' in statistical mechanics}.
\newblock {\em American Journal of Physics}, 83:545--554, June 2015.

\bibitem{Les80}
A.~M. {Lesk}.
\newblock {On the Gibbs paradox - What does indistinguishability really mean}.
\newblock {\em Journal of Physics A Mathematical General}, 13:L111--L114, April
  1980.

\bibitem{Fre14}
D.~{Frenkel}.
\newblock {Why Colloidal Systems Can Be Described by Statistical Mechanics:
  Some Not Very Original Comments on the Gibbs Paradox}.
\newblock {\em Molecular Physics}, 112:2325--2329, September 2014.

\bibitem{C+M15}
M.~E. {Cates} and V.~N. {Manoharan}.
\newblock {Celebrating Soft Matter's 10th anniversary: Testing the foundations
  of classical entropy: colloid experiments}.
\newblock {\em Soft Matter}, 11:6538--6546, 2015.

\bibitem{Kam84}
N.~G. {van Kampen}.
\newblock {The {Gibbs} Paradox}.
\newblock In W.~E. {Parry}, editor, {\em Essays in Theoretical Physics}, page
  303, 1984.

\bibitem{Blumenfeld}
Lev~A. Blumenfeld and Alexander~Y. Grosberg.
\newblock Gibbs paradox and the notion of construction in thermodynamics and
  statistical physics.
\newblock {\em Biofizika (Moscow)}, 40(3):660 -- 667, 1995.

\bibitem{Jay96}
E.~T. {Jaynes}.
\newblock {The Gibbs Paradox}.
\newblock In G.~J. {Erickson}, P.~{Neudorfer}, and C.~R. {Smith}, editors, {\em
  {Maximum-Entropy and Bayesian Methods}}, 1992.

\bibitem{T+C02}
C.-Y. {Tseng} and A.~{Caticha}.
\newblock {Yet another resolution of the Gibbs paradox: an information theory
  approach}.
\newblock In R.~L. {Fry}, editor, {\em Bayesian Inference and Maximum Entropy
  Methods in Science and Engineering}, volume 617 of {\em American Institute of
  Physics Conference Series}, pages 331--339, May 2002.

\bibitem{Fon64}
P.~Fong.
\newblock Semipermeable membrane and the {Gibbs} paradox.
\newblock {\em American Journal of Physics}, 32(2):170--171, 1964.

\bibitem{V+D11}
M.~A.~M. {Versteegh} and D.~{Dieks}.
\newblock {The Gibbs paradox and the distinguishability of identical
  particles}.
\newblock {\em American Journal of Physics}, 79:741--746, July 2011.

\bibitem{Phillips_book}
R.B. Phillips, J.~Kondev, and J.~Theriot.
\newblock {\em Physical Biology of the Cell}.
\newblock Garland Science, 2009.

\bibitem{Bialek_book}
W.S. Bialek.
\newblock {\em Biophysics: Searching for Principles}.
\newblock Princeton University Press, 2012.

\bibitem{HZE17}
J.~M. {Horowitz}, K.~{Zhou}, and J.~L. {England}.
\newblock {Minimum energetic cost to maintain a target nonequilibrium state}.
\newblock {\em \pre}, 95(4):042102, April 2017.

\bibitem{MHL12}
A.~Murugan, D.~A. Huse, and S.~Leibler.
\newblock Speed, dissipation, and error in kinetic proofreading.
\newblock {\em Proceedings of the National Academy of Sciences},
  109(30):12034--12039, 2012.

\bibitem{MHL14}
A.~{Murugan}, D.~A. {Huse}, and S.~{Leibler}.
\newblock {Discriminatory Proofreading Regimes in Nonequilibrium Systems}.
\newblock {\em Physical Review X}, 4(2):021016, April 2014.

\end{thebibliography}
\end{document}